\documentclass[aps,prd,reprint,preprintnumbers,nofootinbib,superscriptaddress,floatfix]{revtex4-1} 

\usepackage[normalem]{ulem}    
\usepackage{caption}
\usepackage{tikz,xcolor}
\usepackage[colorlinks = true,
linkcolor = blue,
urlcolor  = blue,
citecolor = blue,
anchorcolor = blue]{hyperref}

\definecolor{lime}{HTML}{A6CE39}
\DeclareRobustCommand{\orcidicon}{%
	\begin{tikzpicture}
		\draw[lime, fill=lime] (0,0)
		circle [radius=0.16]
		node[white] {{\fontfamily{qag}\selectfont \tiny ID}};
		\draw[white, fill=white] (-0.0625,0.095)
		circle [radius=0.007];
	\end{tikzpicture}
	\hspace{-2mm}
}
\foreach \x in {A, ..., Z}{%
\expandafter\xdef\csname orcid\x\endcsname{\noexpand\href{https://orcid.org/\csname orcidauthor\x\endcsname}{\noexpand\orcidicon}}
}

\usepackage{comment}

\usepackage{amsmath}  
\usepackage{amssymb} 
\usepackage{mathtools} 
\usepackage[utf8]{inputenc}

\usepackage{xcolor}   

\usepackage{graphicx,bm}    
\usepackage{booktabs}    
\usepackage{url}    

\usepackage[normalem]{ulem}

\usepackage{ragged2e}

\begin{document} 

\title{Modern Determination of Pion and Kaon Fragmentation Functions from SIA and High-Precision COMPASS SIDIS Multiplicities}

\collaboration{HAPS Collaboration}

\author{Maryam Soleymaninia \orcidB{}} 
\email{Maryam\_Soleymaninia@ipm.ir}
\affiliation{School of Particles and Accelerators, Institute for Research in Fundamental Sciences (IPM), Tehran, Iran}

\author{Hamzeh Khanpour \orcidE{}}
\email{Hamzeh.Khanpour@cern.ch}
\affiliation{AGH University, Faculty of Physics and Applied Computer Science, Al. Mickiewicza 30, 30-055 Krakow, Poland}
\affiliation{School of Particles and Accelerators, Institute for Research in Fundamental Sciences (IPM), Tehran, Iran}

\author{Hubert Spiesberger \orcidS{}}
\email{spiesber@uni-mainz.de}
\affiliation{PRISMA$^{++}$ Cluster of Excellence, Institut f\"ur Physik, Johannes Gutenberg-Universit\"at Mainz, Mainz, Germany}

\author{Majid Azizi \orcidD{}}
\email{Ma.Azizi@ipm.ir}
\affiliation{School of Particles and Accelerators, Institute for Research in Fundamental Sciences (IPM), Tehran, Iran}

\author{Michael~Klasen \orcidG{}}
\email{Michael.Klasen@uni-muenster.de}
\affiliation{Institut f\"ur Theoretische Physik, Universit\"at M\"unster, \\ Wilhelm-Klemm-Stra\ss{}e 9, 48149 M\"unster, Germany}

\author{Hadi Hashamipour \orcidA{}}
\email{H\_Hashamipour@ipm.ir}
\affiliation{School of Particles and Accelerators, Institute for Research in Fundamental Sciences (IPM), Tehran, Iran}

\date{\today}  

\begin{abstract}  

We present two separate determinations of charged-pion and charged-kaon fragmentation functions (FFs), 
denoted \texttt{HAPS-PiFF1.0} and \texttt{HAPS-KaFF1.0}, at next-to-leading order (NLO) and 
within a next-to-next-to-leading-order (NNLO) perturbative QCD setup. 
The analysis combines single-inclusive electron-positron annihilation (SIA)
data with charge-separated semi-inclusive deep-inelastic-scattering
(SIDIS) multiplicities from HERMES and COMPASS. 
A central goal of this work is to incorporate the modern COMPASS SIDIS input, namely the 
COMPASS 2025 proton-target multiplicities and the COMPASS 2026 revised isoscalar-target multiplicities, 
into a common charged-pion and charged-kaon FF analysis and to assess their role in the resulting flavor separation. 
The revised isoscalar data supersede the earlier COMPASS measurements
used in previous global fits. The charge-separated pion multiplicities provide important
constraints on favored and unfavored light-quark fragmentation, while
the kaon measurements enhance the sensitivity to light-quark, unfavored,
and strange-to-kaon fragmentation channels. In both analyses, the gluon
FF remains indirectly constrained in the present SIA+SIDIS framework and
should be interpreted with appropriate caution.  
The extractions are carried out using the publicly
available \textsc{MontBlanc} framework, and the resulting
\texttt{HAPS-PiFF1.0} and \texttt{HAPS-KaFF1.0} replica sets are provided
in the standard \textsc{LHAPDF} format. 

\end{abstract} 

\maketitle

\section{Introduction}\label{sec:introduction}

Fragmentation functions (FFs) describe the nonperturbative transition of
partons into observed hadrons and constitute an essential ingredient of
QCD factorization for processes with identified particles in the final
state~\cite{Collins:1989gx,Collins:1981uw}. They enter the theoretical description of single-inclusive
electron-positron annihilation (SIA), semi-inclusive deep-inelastic
scattering (SIDIS), and identified-hadron production in hadronic
collisions, providing the connection between perturbatively calculable
partonic cross sections and long-distance hadronization dynamics. A precise knowledge of FFs is
therefore required for quantitative tests of perturbative QCD, for
flavor and spin studies in lepton-nucleon scattering, and for the
phenomenology of present and future facilities, including the
Electron-Ion Collider (EIC)~\cite{AbdulKhalek:2021gbh,Proceedings:2026xrb,Soleymaninia:2025cvi}.

Among light identified hadrons, charged pions and charged kaons play
complementary roles. Charged pions, with valence structures
$\pi^+(u\bar d)$ and $\pi^-(\bar u d)$, provide the most direct
access to favored and unfavored light-quark fragmentation channels. In
particular, the comparison of $\pi^+$ and $\pi^-$ production is
sensitive to different combinations of $u$, $d$, $\bar u$, and
$\bar d$ fragmentation, especially when charge-separated SIDIS
multiplicities are included. Charged kaons, with valence structures
$K^+(u\bar s)$ and $K^-(\bar u s)$, provide additional sensitivity
to the strange sector. The separation of light-quark and
strange-to-kaon fragmentation channels,  and the corresponding
charge-conjugate combinations, is therefore central to a reliable
description of identified-kaon production. In both cases, the relevant
flavor information cannot be fully determined from any single process or
dataset; it requires a consistent analysis combining complementary SIA
and SIDIS measurements~
\cite{Abdolmaleki:2021yjf,deFlorian:2007aj,deFlorian:2017lwf,Bertone:2017tyb,AbdulKhalek:2022laj,deFlorian:2014xna}.

SIA data provide particularly clean constraints on FFs because the
corresponding cross sections are independent of parton distribution
functions (PDFs). Measurements over a wide range of center-of-mass
energies, including low-energy experiments and data taken at the
$Z$ pole, constrain charge-summed and electroweak-weighted combinations
of quark, antiquark, and gluon FFs. Flavor-tagged samples further
improve the sensitivity to heavy-quark fragmentation channels.
Nevertheless, SIA alone has limited ability to separate quark from
antiquark fragmentation and to disentangle the detailed flavor
structure of favored and unfavored channels. This limitation is
particularly relevant for pion light-quark separation and for the
strange sector in kaon production. It motivates the inclusion of SIDIS
multiplicities, where the dependence on the initial-state target, the
hadron charge, and the kinematic variables $x$, $z$, and $Q^2$ provides
additional flavor sensitivity through the convolution of PDFs and FFs~
\cite{Soleymaninia:2026xjq,Bertone:2017tyb,AbdulKhalek:2022laj,Moffat:2021dji}.

In SIDIS, identified-hadron multiplicities are measured as ratios of
semi-inclusive to inclusive DIS cross sections. For charged pions and
charged kaons, the comparison of charge-separated $\pi^+$ and $\pi^-$
multiplicities, and similarly of $K^+$ and $K^-$ multiplicities, together
with the use of proton, deuteron, or isoscalar targets, probes different
PDF-weighted combinations of fragmentation functions. However, the use
of SIDIS data also requires a careful treatment of the kinematic region
included in the fit. At low values of the hard scale $Q$ or the hadronic
invariant mass $W$, effects beyond the leading-twist collinear framework
may become relevant, including hadron-mass and target-mass corrections,
residual finite-$P_{hT}$ effects in the integrated collinear treatment,
and possible contamination from the target-fragmentation region
~\cite{Accardi:2014qda,Guerrero:2015wha,Boglione:2016bph,Boglione:2019nwk}.
For this reason, the choice of the lower SIDIS scale cut is an important
component of the present analysis.

Several determinations of light-hadron FFs have been performed over the
past decades. Early global analyses established the standard
phenomenological framework for identified-hadron fragmentation at
next-to-leading order, using SIA data and, in some cases, additional
constraints from SIDIS or hadron-production measurements~
\cite{Binnewies:1994ju,Binnewies:1995kg,Kretzer:2000yf,Kniehl:2000fe,Bourhis:2000gs,Albino:2005me,Albino:2008fy,Hirai:2007cx,deFlorian:2007aj}.
Dedicated DSS analyses provided widely used pion and kaon FF
determinations~\cite{deFlorian:2007aj}. A subsequent DSS update
revisited the pion FFs using a modern global NLO analysis of SIA,
SIDIS, and hadronic-collision data~\cite{deFlorian:2014xna}, while
the corresponding parton-to-kaon fragmentation functions were later
updated in Ref.~\cite{deFlorian:2017lwf}. 
The NNFF1.0
determination introduced a neural-network extraction of charged-pion,
charged-kaon, and proton/antiproton FFs from SIA data at LO, NLO, and
NNLO, with Monte Carlo uncertainties and closure-test validation~
\cite{Bertone:2017tyb}. More recently, the MAP Collaboration performed a
determination of pion and kaon FFs at NLO and NNLO using SIA and SIDIS
data within the \textsc{MontBlanc} framework~
\cite{AbdulKhalek:2022laj,Khalek:2021gxf,MontBlancCode}. Simultaneous PDF/FF analyses
and recent global studies of light charged-hadron fragmentation have
also emphasized the interplay between FFs, PDFs, SIDIS data, and
hadron-collider measurements~\cite{Moffat:2021dji,Gao:2024nkz,Gao:2024dbv,Gao:2025hlm}.

The perturbative description of identified-hadron production has also
advanced substantially. Timelike DGLAP evolution is known through NNLO
accuracy, including both nonsinglet and singlet splitting functions~
\cite{Mitov:2006ic,Moch:2007tx,Almasy:2011eq,Bertone:2015cwa}. The
coefficient functions for SIA have long been available at the
corresponding perturbative orders~
\cite{Altarelli:1979kv,Rijken:1996ns,Rijken:1996vr,Rijken:1996npa}.
For SIDIS, approximate NNLO corrections were used in earlier
phenomenological studies~\cite{Abele:2021nyo}, while recent calculations
have provided exact NNLO QCD corrections to semi-inclusive DIS
coefficient functions~
\cite{Bonino:2024qbh,Goyal:2023zdi,Goyal:2024emo,Bonino:2025qta}.

In the present analysis, the NNLO SIDIS predictions are based on the approximate NNLO 
coefficient functions. Accordingly, throughout this work, the label "NNLO" denotes exact 
NNLO timelike evolution and exact NNLO SIA coefficient functions, 
combined with approximate NNLO SIDIS corrections. 
These theoretical developments make it timely to compare charged-pion
and charged-kaon FF extractions at NLO and NNLO in a common framework.
 
Such a comparison provides a controlled test of the perturbative
stability of the fits upon the inclusion of higher-order QCD
corrections, although it should not be interpreted as a complete
estimate of theory uncertainty.  
The theoretical framework used in this analysis follows the
collinear-factorization formalism for SIA and SIDIS observables
implemented in the \textsc{MontBlanc} fitting framework~\cite{MontBlancCode,MontBlancCodezenodo}.

A central motivation for the present work is the recent update of the
COMPASS SIDIS multiplicity data. The COMPASS Collaboration has published
proton-target multiplicities for positive and negative pions, kaons, and
unidentified charged hadrons from deep-inelastic scattering of muons off
a liquid hydrogen target~\cite{COMPASS:2024gje}. Subsequently, COMPASS
provided revised isoscalar-target multiplicities in a dedicated
addendum, using an updated treatment of radiative corrections based on
\textsc{DJANGOH}~\cite{COMPASS:2025bfn}. These revised isoscalar
multiplicities supersede the earlier COMPASS isoscalar measurements of
charged pions, charged kaons, and unidentified charged hadrons~
\cite{COMPASS:2016xvm,COMPASS:2016crr}. It is therefore important, in a
modern global FF analysis, to combine the COMPASS 2025 proton-target
data with the revised COMPASS 2026 isoscalar-target multiplicities,
rather than with the superseded 2017 isoscalar data.

In this paper we present a combined study of charged-pion and charged-kaon FFs, 
denoted \texttt{HAPS-PiFF1.0} and \texttt{HAPS-KaFF1.0}, respectively. 
A primary goal of the present analysis is to incorporate the modern
COMPASS SIDIS multiplicities into charged-pion and charged-kaon FF
determinations and to examine the resulting flavor structure within a
common SIA+SIDIS framework. 
In particular, we include both the COMPASS 
proton-target $\pi^\pm$ and $K^\pm$ multiplicities published in 2025 and 
the revised COMPASS isoscalar-target $\pi^\pm$ and $K^\pm$ multiplicities published in 2026, 
which supersede the earlier COMPASS isoscalar measurements used in previous global analyses. 
The analysis combines these data with SIA measurements and charge-separated SIDIS multiplicities 
from HERMES within a common SIA+SIDIS framework.

The charged-pion and charged-kaon determinations are performed at NLO and NNLO using 
the same kinematic selections within each hadron analysis, allowing us to assess the 
perturbative stability of the extracted FFs and the impact of the updated 
COMPASS input on their flavor decomposition. The extraction is carried out using the publicly 
available \textsc{MontBlanc} framework, with modifications appropriate to the updated charged-pion 
and charged-kaon datasets and to the \texttt{HAPS-PiFF1.0} and \texttt{HAPS-KaFF1.0} setups. 
The uncertainties are estimated using Monte Carlo replicas, and the resulting FF sets are provided 
in the standard \textsc{LHAPDF} format~\cite{Buckley:2014ana,HAPS-PiFF10,HAPS-KaFF10}. 
The present analysis is complementary to the recent \texttt{HAPS-hFF1.0} determination of 
unidentified charged-hadron FFs, which also used the modern COMPASS SIDIS input to 
reassess light charged-hadron fragmentation~\cite{Soleymaninia:2026xjq}.

The paper is organized as follows. In Sec.~\ref{sec:Theoretical} we
summarize the theoretical framework, including collinear factorization
for SIA and SIDIS, timelike DGLAP evolution, the perturbative setup, the
flavor bases, and the FF parametrization. In Sec.~\ref{sec:data} we
describe the experimental datasets, the treatment of the modern COMPASS
SIDIS multiplicities, and the kinematic cuts.  
The main results are presented in Sec.~\ref{sec:results},
including the dataset-level fit quality, the description of COMPASS
multiplicities, and the \texttt{HAPS-PiFF1.0} and \texttt{HAPS-KaFF1.0} FF
sets. Finally, Sec.~\ref{sec:Summary} summarizes our conclusions.

\section{Theoretical framework}\label{sec:Theoretical}

The formalism used in the present analysis follows the standard
collinear-factorization description of identified-hadron production in
perturbative QCD. The same theoretical framework is used for the
charged-pion and charged-kaon determinations, denoted
\texttt{HAPS-PiFF1.0} and \texttt{HAPS-KaFF1.0}, respectively. In each
case, a common set of FFs is used to describe single-inclusive
electron-positron annihilation (SIA) and semi-inclusive
deep-inelastic-scattering (SIDIS) multiplicities for the corresponding
identified hadron. The observed hadron is therefore
$h=\pi^\pm$ or $h=K^\pm$, depending on the dataset and on the FF set
under consideration. The FFs are parametrized at an input scale $Q_0$
and evolved to the scales of the data using timelike DGLAP evolution.
Both analyses are performed at next-to-leading order (NLO) and
next-to-next-to-leading order (NNLO), allowing us to assess the
perturbative stability of the extractions under the inclusion of
higher-order QCD corrections. The theoretical framework follows the
factorization properties of hard processes in QCD~\cite{Collins:1989gx,Collins:1981uw}, 
together with the perturbative description of SIA and 
SIDIS coefficient functions~\cite{Altarelli:1979kv,Graudenz:1994dq}.
In this analysis, the calculation is performed within the 
standard collinear-factorization
framework for SIA and SIDIS observables, following the implementation
provided by the \textsc{MontBlanc} fitting framework~\cite{MontBlancCode}.

\subsection{Collinear factorization}

In collinear factorization, the long-distance transition of a parton
$i$ into an observed hadron $h$ is described by a fragmentation function
$D_i^h(z,\mu_F)$, where $z$ denotes the fraction of the parton momentum
carried by the detected hadron and $\mu_F$ is the final-state
factorization scale. In the present work, the identified hadron is a
charged pion or a charged kaon, $h=\pi^\pm,K^\pm$, with
charge-separated predictions used whenever required by the experimental
multiplicities. The FFs are nonperturbative objects and must be
determined from data, while their scale dependence is governed by
perturbative QCD evolution.

For observables involving an identified hadron in the final state, the
cross section can be written schematically as a convolution of
perturbatively calculable partonic coefficient functions with
nonperturbative FFs. For processes such as SIA, which do not involve a
hadronic initial state, the factorized structure may be written as
%
\begin{equation}
d\sigma^h =
\sum_i d\hat{\sigma}_i \otimes D_i^h ,
\label{eq:fact_sia_schematic}
\end{equation}
%
where $d\hat{\sigma}_i$ denotes the short-distance cross section for
producing a parton of flavor $i$, and the symbol $\otimes$ denotes the
appropriate convolution over partonic momentum fractions. For SIDIS
multiplicities on a hadronic target $N$, the cross section also depends
on the PDFs of the target and has the schematic form
%
\begin{equation}
d\sigma_N^h =
\sum_{i,j}
f_i^N \otimes d\hat{\sigma}_{ij} \otimes D_j^h .
\label{eq:fact_sidis_schematic}
\end{equation}
%
Here $f_i^N(x,\mu_F)$ is the PDF of parton flavor $i$ in the target
$N$, $d\hat{\sigma}_{ij}$ is the perturbative hard-scattering
coefficient, and $D_j^h(z,\mu_F)$ is the FF for the parton $j$
fragmenting into the observed hadron $h$.

The factorization scale $\mu_F$ separates the perturbatively calculable
short-distance dynamics from the nonperturbative PDFs and FFs, while the
renormalization scale $\mu_R$ enters through the running strong coupling
$\alpha_s(\mu_R)$. In SIA the theoretical prediction depends only on FFs
and perturbative coefficient functions, whereas in SIDIS it depends
simultaneously on PDFs, FFs, and the inclusive DIS structure functions
entering the multiplicity denominator. This complementarity is essential
for both charged-pion and charged-kaon FFs. SIA provides clean
constraints on charge-summed and electroweak-weighted combinations,
while SIDIS adds charge- and target-dependent flavor sensitivity that is
not directly available from inclusive SIA data alone.

\subsection{Single-inclusive annihilation}

SIA provides one of the cleanest constraints on fragmentation functions,
since the process $e^+e^-\to hX$ is free of PDF uncertainties. For
charged-pion and charged-kaon production, the measured observables are
typically normalized single-inclusive spectra for $h=\pi^\pm$ or
$h=K^\pm$, or, depending on the experimental convention, the
corresponding charge-summed combinations $\pi^+ + \pi^-$ and
$K^+ + K^-$. A representative normalized SIA observable may be written
as
%
\begin{equation}
F^h(z,Q^2)
\equiv
\frac{1}{\sigma_{\rm tot}}
\frac{d\sigma^h}{dz},
\label{eq:sia_normalized}
\end{equation}
%
where $Q=\sqrt{s}$ is the center-of-mass energy and $z$ is the
experimental scaling variable, usually related to the hadron energy or
scaled momentum. In the numerical implementation, the definition of $z$
and the normalization are taken consistently from the corresponding
experimental data tables.

The factorized expression for the SIA observable can be written
schematically as
%
\begin{equation}
F^h(z,Q^2)
=
\sum_i
C_i^{\rm SIA}
\left(
z,\alpha_s(\mu_R),
\frac{Q^2}{\mu_F^2},
\frac{Q^2}{\mu_R^2}
\right)
\otimes
D_i^h(z,\mu_F),
\label{eq:sia_factorized}
\end{equation}
%
where $C_i^{\rm SIA}$ are perturbatively calculable coefficient
functions and the sum runs over quarks, antiquarks, and the gluon. The
coefficient functions are included consistently at the perturbative
order of the fit. The NLO and NNLO SIA coefficient functions are known
and form an essential component of modern FF extractions~\cite{Rijken:1996ns,Rijken:1996vr,Rijken:1996npa}.

Inclusive SIA measurements constrain charge-summed and electroweak
charge-weighted combinations of FFs. Data taken at different
center-of-mass energies provide information on scaling violations, while
flavor-tagged measurements at the $Z$ pole improve sensitivity to
heavy-quark fragmentation. However, SIA alone has limited ability to
separate quark from antiquark fragmentation and to distinguish favored
from unfavored channels. This limitation is relevant for the separation
of light-quark pion FFs and is particularly important for kaons, where
the strange-to-kaon channels require additional flavor information. The
present analysis therefore combines SIA data with charge-separated SIDIS
multiplicities, following the strategy used in modern global
identified-hadron FF determinations~\cite{Khalek:2021gxf,AbdulKhalek:2022laj}.

For charge-summed SIA data, the relevant FF combinations are
$D_i^{\pi^+}+D_i^{\pi^-}$ or $D_i^{K^+}+D_i^{K^-}$, depending on the
hadron species. Charge-separated information is provided primarily by
SIDIS. The treatment of the individual SIA datasets, including their
normalization conventions, flavor tags, point counts, and kinematic
cuts, is described in Sec.~\ref{sec:data}.

\subsection{Semi-inclusive DIS multiplicities}

SIDIS multiplicities provide complementary information to SIA by
combining final-state hadron identification with the flavor structure of
the initial-state nucleon. For the processes
$\ell N\to \ell'\pi^\pm X$ and $\ell N\to \ell'K^\pm X$, the
identified-hadron multiplicity is defined as the ratio of the
semi-inclusive DIS cross section to the inclusive DIS cross section. In
the notation used here,
%
\begin{equation}
M_N^h(x,z,Q^2) =
\frac{
d\sigma_N^h/(dx\,dQ^2\,dz)
}{
d\sigma_N^{\rm DIS}/(dx\,dQ^2)
};\,\,\,
h=\pi^+,\pi^-,K^+,K^- .
\label{eq:sidis_multiplicity}
\end{equation}
%
The target $N$ denotes the proton, deuteron, or isoscalar target,
depending on the experimental dataset. In practice, the experimental
multiplicities are provided in finite bins of $x$, $y$, and $z$, and
the theoretical predictions must be evaluated consistently with the
binning and kinematic cuts of the corresponding measurement.

The semi-inclusive numerator has the schematic factorized form
%
\begin{equation}
d\sigma_N^h
=
\sum_{i,j}
f_i^N(x,\mu_F)
\otimes
d\hat{\sigma}_{ij}^{\rm SIDIS}
\otimes
D_j^h(z,\mu_F),
\label{eq:sidis_numerator}
\end{equation}
%
while the inclusive DIS denominator is written as
%
\begin{equation}
d\sigma_N^{\rm DIS}
=
\sum_i
f_i^N(x,\mu_F)
\otimes
d\hat{\sigma}_{i}^{\rm DIS}.
\label{eq:sidis_denominator}
\end{equation}
%
Here $d\hat{\sigma}_{ij}^{\rm SIDIS}$ and
$d\hat{\sigma}_{i}^{\rm DIS}$ denote the perturbative coefficient
functions entering the semi-inclusive and inclusive cross sections,
respectively. The SIDIS coefficient functions are included at the
perturbative order of the fit. The NLO formalism has been known for a
long time~\cite{Graudenz:1994dq}, while approximate and exact NNLO
corrections have recently become available~\cite{Abele:2021nyo,Bonino:2024qbh,Goyal:2023zdi,Goyal:2024emo,Bonino:2025qta}.
In the present analysis, the numerical implementation of 
SIDIS observables follows the setup provided by the
\textsc{MontBlanc} fitting framework~\cite{MontBlancCode}.

The flavor sensitivity of SIDIS arises from the simultaneous dependence
on PDFs and FFs. For a proton target, the larger $u$-quark density
enhances sensitivity to fragmentation channels weighted by the
corresponding proton PDFs, while an isoscalar target provides a more
balanced weighting of light flavors. The comparison of $\pi^+$ and
$\pi^-$ multiplicities primarily constrains favored and unfavored
light-quark fragmentation into charged pions. The comparison of $K^+$
and $K^-$ multiplicities provides additional sensitivity to
strange-to-kaon fragmentation, although the strange sector remains
correlated with the PDFs, unfavored FFs, and the chosen flavor
parametrization.

The \texttt{HAPS-PiFF1.0} and \texttt{HAPS-KaFF1.0} fits include the
corresponding modern COMPASS charged-pion and charged-kaon
multiplicities. These consist of the COMPASS 2025 proton-target
$\pi^\pm$ and $K^\pm$ data~\cite{COMPASS:2024gje} and the COMPASS 2026 revised
isoscalar-target $\pi^\pm$ and $K^\pm$ data~\cite{COMPASS:2025bfn}, together with the HERMES
charged-pion and charged-kaon multiplicities included in the final
datasets~\cite{HERMES:2012uyd}. The revised
COMPASS isoscalar multiplicities supersede the older COMPASS isoscalar
charged-pion, charged-kaon, and unidentified charged-hadron
measurements
\cite{COMPASS:2016xvm,COMPASS:2016crr}. A detailed discussion of these
datasets, their point counts, and their kinematic coverage is given in
Sec.~\ref{sec:data}.

The use of SIDIS data also requires a careful assessment of the
kinematic region in which a leading-twist collinear description is
expected to be reliable. At low values of $Q$, low final-state invariant
mass, or extreme values of $z$, effects such as hadron-mass corrections and target-fragmentation contributions
may become relevant~\cite{Accardi:2014qda,Guerrero:2015wha,Boglione:2016bph,Boglione:2019nwk}.
For this reason, the lower SIDIS scale cut is treated as an important
part of the fit definition. The stability of the description under
variations of this cut is examined in Sec.~\ref{Kinematic_cuts}.

\subsection{Timelike DGLAP evolution and perturbative order}

The scale dependence of FFs is governed by the timelike DGLAP evolution
equations. For a hadron $h$, the evolution of the FF $D_i^h(z,\mu^2)$ is
given by
%
\begin{equation}
\frac{dD_i^h(z,\mu^2)}{d\ln\mu^2}
=
\sum_j
P_{ij}^{T}(z,\alpha_s(\mu^2))
\otimes
D_j^h(z,\mu^2),
\label{eq:timelike_dglap}
\end{equation}
%
where $P_{ij}^{T}$ are the timelike splitting functions and the
convolution is performed in the momentum fraction $z$. The perturbative
expansion of the splitting functions is written as
%
\begin{equation}
P_{ij}^{T}(z,\alpha_s)
=
\sum_{n=0}^{N}
\left(\frac{\alpha_s}{2\pi}\right)^{n+1}
P_{ij}^{T,(n)}(z),
\label{eq:splitting_expansion}
\end{equation}
%
with $N=1$ for NLO evolution and $N=2$ for NNLO evolution. The NNLO
timelike splitting functions are known for both nonsinglet and singlet
sectors~~\cite{Mitov:2006ic,Moch:2007tx,Almasy:2011eq,Bertone:2015cwa,Chen:2020uvt} 
and have been benchmarked in modern evolution 
frameworks such as APFEL and APFEL++~\cite{Bertone:2013vaa,Bertone:2017gds}.

The perturbative coefficient functions entering SIA and SIDIS are
expanded in the same way,
%
\begin{equation}
C_i =
\sum_{n=0}^{N}
\left(\frac{\alpha_s}{2\pi}\right)^n
C_i^{(n)} ,
\label{eq:coeff_expansion}
\end{equation}
%
where the upper limit $N$ is chosen consistently with the perturbative
order of the fit. 
At NLO, the analysis employs NLO timelike splitting functions and NLO coefficient 
functions for both SIA and SIDIS. At NNLO, the FFs are evolved using the exact 
NNLO timelike splitting functions, and the SIA observables are evaluated with the 
exact NNLO SIA coefficient functions. For SIDIS, however, we use the approximate 
NNLO coefficient functions of Ref.~\cite{Abele:2021nyo}, based on the dominant 
partonic-threshold contributions, as implemented in the MontBlanc framework.

%
%
%
%
%
The comparison between NLO and NNLO is therefore interpreted as a
test of perturbative stability within the fixed-order,
massless collinear-factorization setup adopted in this analysis. 
The input FFs are parametrized at the scale
$Q_0=5~{\rm GeV}$ and evolved to the scales of the experimental data
using the timelike DGLAP equations implemented in
\textsc{APFEL++}~\cite{Bertone:2017gds}. The strong coupling is fixed
at $\alpha_s(M_Z)=0.118,$
and the charm- and bottom-flavor thresholds are set to
%
\begin{equation}
\mu_c=1.51~{\rm GeV},
\qquad
\mu_b=4.92~{\rm GeV},
\label{eq:alphas_thresholds}
\end{equation}
%
respectively. These quantities are referred to as flavor thresholds.
	
For the SIDIS multiplicities, the required collinear PDFs are taken
from the perturbative-charm variants of NNPDF4.0 at the corresponding
perturbative order~\cite{NNPDF:2021njg}. Specifically, the NLO fits use
the \texttt{NNPDF40\_nlo\_pch\_as\_01180} set, while the NNLO fits use
the corresponding NNPDF4.0 NNLO perturbative-charm set. The PDFs
therefore undergo spacelike evolution as provided by NNPDF4.0, whereas
the FFs undergo timelike evolution through \textsc{APFEL++}. The same
general perturbative prescription is adopted for the charged-pion and
charged-kaon determinations, while their fitted datasets, kinematic
cuts, and flavor bases are hadron-specific. All SIA and SIDIS
predictions are evaluated consistently within the
\textsc{MontBlanc} fitting framework~\cite{MontBlancCode}.

The dependence on the proton PDFs is expected to be limited for SIDIS
multiplicities because PDFs enter both the semi-inclusive numerator and the
inclusive DIS denominator, leading to a partial cancellation
\cite{Khalek:2021gxf}.

\subsection{Flavor basis and charge conjugation}

The charged-pion and charged-kaon analyses require explicit definitions
of the flavor basis and charge-conjugation relations. In each case, the
independent FFs are defined for the positively charged hadron,
$\pi^+$ or $K^+$, and the corresponding negatively charged FFs are
obtained by charge conjugation. The flavor basis is therefore hadron-dependent and must be
specified separately for pions and kaons.

\subsubsection{Charged pions}

For charged pions, the valence structures are
$\pi^+\sim u\bar d$ and $\pi^-\sim \bar u d$. Charge conjugation gives
\begin{equation} 
\begin{aligned} D_q^{\pi^-}(z,Q) &= D_{\bar q}^{\pi^+}(z,Q),\\ D_{\bar q}^{\pi^-}(z,Q) &= D_q^{\pi^+}(z,Q),\\ D_g^{\pi^-}(z,Q) &= D_g^{\pi^+}(z,Q). \end{aligned} 
\label{eq:charge_conjugation_pion} 
\end{equation}
In this QCD analysis, the independent charged-pion basis is
\begin{equation} 
\begin{aligned} & D_u^{\pi^+},\, D_{\bar d}^{\pi^+},\, D_d^{\pi^+}=D_{\bar u}^{\pi^+},\\ & D_s^{\pi^+}=D_{\bar s}^{\pi^+},\, D_c^{\pi^+}=D_{\bar c}^{\pi^+},\\ & D_b^{\pi^+}=D_{\bar b}^{\pi^+},\, D_g^{\pi^+} . 
\end{aligned} 
\label{eq:pion_positive_basis} 
\end{equation}
No additional isospin relation between the two favored pion FFs is
imposed; in particular, $D_u^{\pi^+}$ and $D_{\bar d}^{\pi^+}$ are kept
as independent distributions in this basis.

Using Eq.~\eqref{eq:charge_conjugation_pion}, the corresponding
$\pi^-$ flavor map is
\begin{align} 
D_u^{\pi^-} &=D_{\bar u}^{\pi^+} =D_d^{\pi^+}, & D_{\bar u}^{\pi^-} &=D_u^{\pi^+}, \nonumber\\ D_d^{\pi^-} &=D_{\bar d}^{\pi^+}, & D_{\bar d}^{\pi^-} &=D_d^{\pi^+} =D_{\bar u}^{\pi^+}, \nonumber\\ D_s^{\pi^-} &=D_{\bar s}^{\pi^-} =D_s^{\pi^+} =D_{\bar s}^{\pi^+}, & D_g^{\pi^-} &=D_g^{\pi^+}, \nonumber\\ D_Q^{\pi^-} &=D_{\bar Q}^{\pi^-} =D_Q^{\pi^+} =D_{\bar Q}^{\pi^+}, & Q&=c,b . \label{eq:pion_negative_map} 
\end{align}
In the valence-quark sense, $D_u^{\pi^+}$ and
$D_{\bar d}^{\pi^+}$ are favored for $\pi^+$ production, while
$D_{\bar u}^{\pi^+}=D_d^{\pi^+}$ and the strange and heavy-flavor
channels are unfavored or sea-like. The favored channels for $\pi^-$
follow by charge conjugation.

\subsubsection{Charged kaons}

For charged kaons, the valence structures are
$K^+\sim u\bar s$ and $K^-\sim \bar u s$. Charge conjugation gives
\begin{equation}
\begin{aligned}
D_q^{K^-}(z,Q) &=
D_{\bar q}^{K^+}(z,Q),\\
D_{\bar q}^{K^-}(z,Q) &=
D_q^{K^+}(z,Q),\\
D_g^{K^-}(z,Q) &=
D_g^{K^+}(z,Q).
\end{aligned}
\label{eq:charge_conjugation_kaon}
\end{equation}
For the charged-kaon fit we adopt a reduced flavor basis for $K^+$ in which the 
favored channels $D_u^{K^+}$ and $D_{\bar{s}}^{K^+}$ are independent, while the 
unfavored light-quark sector is constrained by $D_s^{K^+}=D_{\bar{u}}^{K^+}$ and $D_d^{K^+}=D_{\bar{d}}^{K^+}$. 
These equalities are fit-basis constraints rather than consequences of charge conjugation. 
The independent charged-kaon basis is therefore: 
\begin{equation}
\begin{aligned}
&
D_u^{K^+},\,
D_{\bar s}^{K^+},\,
D_s^{K^+}=D_{\bar u}^{K^+},\\
&
D_d^{K^+}=D_{\bar d}^{K^+},\,
D_c^{K^+}=D_{\bar c}^{K^+},\\
&
D_b^{K^+}=D_{\bar b}^{K^+},\,
D_g^{K^+}.
\end{aligned}
\label{eq:kaon_positive_basis}
\end{equation}
This basis keeps the favored strange-antiquark FF
$D_{\bar s}^{K^+}$ independent from $D_s^{K^+}$. Therefore
$D_s^{K^+}$ and $D_{\bar s}^{K^+}$ should not be assumed equal unless
such a constraint is explicitly imposed.

We have also examined the sensitivity of the strange-sector
determination to the reduced-basis assumption
$D_s^{K^+}=D_{\bar u}^{K^+}$ by performing a dedicated variant fit in
which these two FFs are parametrized independently, while all other
ingredients of the analysis are kept unchanged. The resulting changes
in the fit quality and in the extracted $D_s^{K^+}$,
$D_{\bar u}^{K^+}$, and $D_{\bar s}^{K^+}$ distributions are discussed
in detail in Sec.~\ref{subsec:kaon-basis-stability}.

Using Eq.~\eqref{eq:charge_conjugation_kaon}, the corresponding
$K^-$ flavor map is
\begin{align}
D_u^{K^-}
&=D_{\bar u}^{K^+}
 =D_s^{K^+},
&
D_{\bar u}^{K^-}
&=D_u^{K^+},
\nonumber\\
D_s^{K^-}
&=D_{\bar s}^{K^+},
&
D_{\bar s}^{K^-}
&=D_s^{K^+}
 =D_{\bar u}^{K^+},
\nonumber\\
D_d^{K^-}
&=D_{\bar d}^{K^+}=D_d^{K^+}
 =D_{\bar d}^{K^-},
& 
\nonumber\\
D_Q^{K^-}
&=D_{\bar Q}^{K^-},
\quad
\nonumber\\
D_Q^{K^-}
&=D_Q^{K^+}
 =D_{\bar Q}^{K^+},
\quad Q=c,b,
\nonumber\\
D_g^{K^-}
&=D_g^{K^+}.
\label{eq:kaon_negative_map}
\end{align}
In the valence-quark sense, $D_u^{K^+}$ and
$D_{\bar s}^{K^+}$ are favored for $K^+$ production. The channels
$D_{\bar u}^{K^+}=D_s^{K^+}$ and
$D_d^{K^+}=D_{\bar d}^{K^+}$ are unfavored.
For $K^-$ production, the favored channels are obtained by charge
conjugation, namely $D_{\bar u}^{K^-}$ and $D_s^{K^-}$.

When quark-plus combinations are displayed, we use the convention
\begin{equation}
\begin{aligned}
D_{q^+}^{h}(z,Q)
&\equiv
D_q^{h}(z,Q)+D_{\bar q}^{h}(z,Q),
\\
h&=\pi^+,\,\pi^-,\,K^+,\,K^- .
\end{aligned}
\label{eq:qplus_definition}
\end{equation}
These $q^+$ combinations are displayed combinations, not additional
independent FFs. For example, if the fitted basis imposes
$D_c^h=D_{\bar c}^h$ or $D_b^h=D_{\bar b}^h$, then
$D_{c^+}^h=2D_c^h$ and $D_{b^+}^h=2D_b^h$. 
The equalities imposed in Eqs.~(\ref{eq:pion_positive_basis}) and
(\ref{eq:kaon_positive_basis}) define the reduced flavor basis adopted in
the fit and should be understood as fit-basis assumptions. 
Previous charged-pion and charged-kaon FF analyses provide
useful benchmarks for these flavor-basis choices~\cite{Abdolmaleki:2021yjf, Moffat:2021dji,deFlorian:2007aj,deFlorian:2017lwf,Bertone:2017tyb,AbdulKhalek:2022laj}.

\subsection{Parametrization and uncertainty propagation}

The nonperturbative input FFs are parametrized at the initial scale
$Q_0=5~{\rm GeV}$ using a flexible neural-network representation.
Following the strategy of modern neural-network FF determinations, the
parametrization is chosen to reduce functional bias while allowing the
data to constrain the shape of the FFs over the fitted $z$ range~\cite{Bertone:2017tyb,AbdulKhalek:2022laj}. 
The parametrization is
defined independently for the positively charged hadron in each
analysis, $h=\pi^+$ or $K^+$, and the corresponding negatively charged
FFs are obtained by charge conjugation.
\begin{equation}
zD_i^h(z,Q_0) = \left[
N_i^h(z;\boldsymbol{\theta})-
N_i^h(1;\boldsymbol{\theta})
\right]^2,
\qquad h=\pi^+,K^+ ,
\label{eq:nn_parametrization}
\end{equation}
where $N_i^h(z;\boldsymbol{\theta})$ denotes the neural-network output
for flavor combination $i$, and $\boldsymbol{\theta}$ is the set of neural network weights and biases. 
The subtraction of the network output at $z=1$ imposes the
endpoint condition
\begin{equation}
D_i^h(z=1,Q_0)=0,
\end{equation}
while the squared output enforces positivity of the input FFs at
$Q_0$. No additional power-like preprocessing factor is introduced in
the baseline parametrization. 
For both \texttt{HAPS-PiFF1.0} and \texttt{HAPS-KaFF1.0}, the input FFs are represented by 
a single one-hidden-layer feed-forward neural network. 
The network has one input node, corresponding to the momentum fraction $z$, 20 hidden nodes 
with sigmoid activation functions, and seven output nodes with linear activation functions. 
The seven outputs correspond to the independent FF combinations defined for the positively 
charged hadron, $\pi^+$ or $K^+$, in the pion and kaon flavor bases, respectively. 
With this architecture, denoted $[1,20,7]$, each hadron-specific fit 
contains 187 neural-network parameters before minimization. 

To control overfitting, we use look-back cross-validation. For each Monte
Carlo replica, datasets containing more than ten points are randomly
divided into equally sized training and validation subsets, whereas
smaller datasets are assigned entirely to training. Only the training
subset enters the minimization, while the validation $\chi^2$ is monitored
independently. The parameters corresponding to the absolute minimum of the
validation $\chi^2$ along the optimization trajectory are retained.
Replicas with $\chi^2_{\rm tot}/N_{\rm dat}>3$ are subsequently discarded.
The final ensembles at each perturbative order contain
$N_{\rm rep}=200$ accepted replicas for both pions and kaons.
When more than 200 replicas satisfy the post-fit criterion, the standard MontBlanc grid-building 
procedure orders the surviving fits according to their global \(\chi^2\) and retains 
the 200 replicas with the smallest values.

Experimental uncertainties are propagated to the FFs using a Monte Carlo
replica method. The covariance matrix includes both uncorrelated and
correlated sources of experimental uncertainty, and the data replicas are
generated from this covariance matrix following the MAPFF1.0 Monte Carlo
sampling procedure described in Ref.~\cite{Khalek:2021gxf}. For each
data replica, the fit is repeated and a corresponding FF replica is
obtained. The central value is then computed as the average over the
ensemble,
%
\begin{equation}
\left\langle D_i^h(z,Q)\right\rangle
=
\frac{1}{N_{\rm rep}}
\sum_{k=1}^{N_{\rm rep}}
D_{i,k}^h(z,Q),
\label{eq:replica_mean}
\end{equation}
%
where $N_{\rm rep}$ is the number of fitted replicas. The
one-standard-deviation uncertainty is estimated as
%
\begin{equation}
\Delta D_i^h(z,Q)
=
\left[
\frac{1}{N_{\rm rep}-1}
\sum_{k=1}^{N_{\rm rep}}
\left(
D_{i,k}^h(z,Q)
-
\left\langle D_i^h(z,Q)\right\rangle
\right)^2
\right]^{1/2}.
\label{eq:replica_std}
\end{equation}

PDF uncertainties are propagated by pairing each data replica with PDF replicas from the 
corresponding \texttt{NNPDF4.0} ensemble. The quoted FF uncertainty therefore includes 
both experimental and PDF-induced components within the adopted replica procedure.  
It does not, by itself, include all possible sources of theory uncertainty unless additional 
variations, such as scale, PDF-set, kinematic-cut, or parametrization variations, are explicitly included.

The minimization is performed following the strategy implemented in the
\textsc{MontBlanc} fitting framework~\cite{MontBlancCode,MontBlancCodezenodo}. The
optimization is carried out with the Ceres Solver~\cite{Agarwal:Ceres},
while analytic derivatives of the feed-forward neural-network
parametrization are evaluated using the NNAD code
~\cite{AbdulKhalek:2020uza}. This setup provides an efficient and
reproducible minimization procedure for the Monte Carlo replica fits.
After minimization, the resulting \texttt{HAPS-PiFF1.0} and
\texttt{HAPS-KaFF1.0} FF replicas~\cite{HAPS-PiFF10,HAPS-KaFF10} are exported in the standard
\textsc{LHAPDF} format~~\cite{Buckley:2014ana} for phenomenological applications.

It should be noted that the pion and kaon fits use members drawn from the same NNPDF4.0 ensemble,
but their PDF-member assignments are not aligned: pion replica $k$ and kaon
replica $k$ generally correspond to different input PDF members. The two
released ensembles should therefore not be paired according to their replica
indices. They provide separate marginal FF uncertainties but do not encode
the PDF-induced pion-kaon cross-covariance required for a correlated
replica-by-replica combination. Such a combination would require aligned
PDF-member assignments or a simultaneous pion--kaon analysis.

\section{Experimental datasets and kinematic cuts}\label{sec:data}

The \texttt{HAPS-PiFF1.0} and \texttt{HAPS-KaFF1.0} analyses are based
on charge-separated charged-pion and charged-kaon data from SIA and
SIDIS. The SIA measurements provide clean constraints on charge-summed
and electroweak-weighted combinations of fragmentation functions, while
the SIDIS multiplicities provide additional flavor sensitivity through
their dependence on the target, the hadron charge, and the DIS
kinematics. The two determinations use the same general fitting
framework and kinematic selection strategy, but the final datasets are
hadron-specific. 

After the kinematic selections described below, the charged-pion dataset
contains $N_{\rm dat}^{\pi}=1295$ points, of which $377$ are from SIA
measurements and $918$ are from SIDIS multiplicities. 
The charged-kaon dataset contains $N_{\rm dat}^{K}=1134$ points, of which
$339$ are from SIA measurements and $795$ are from SIDIS
multiplicities. 
The corresponding dataset-by-dataset composition and fit quality are summarized in
Table~\ref{tab:haps-kaff-piff-chi2}.

\begin{figure*}[t]
\centering
\includegraphics[width=0.49\textwidth]{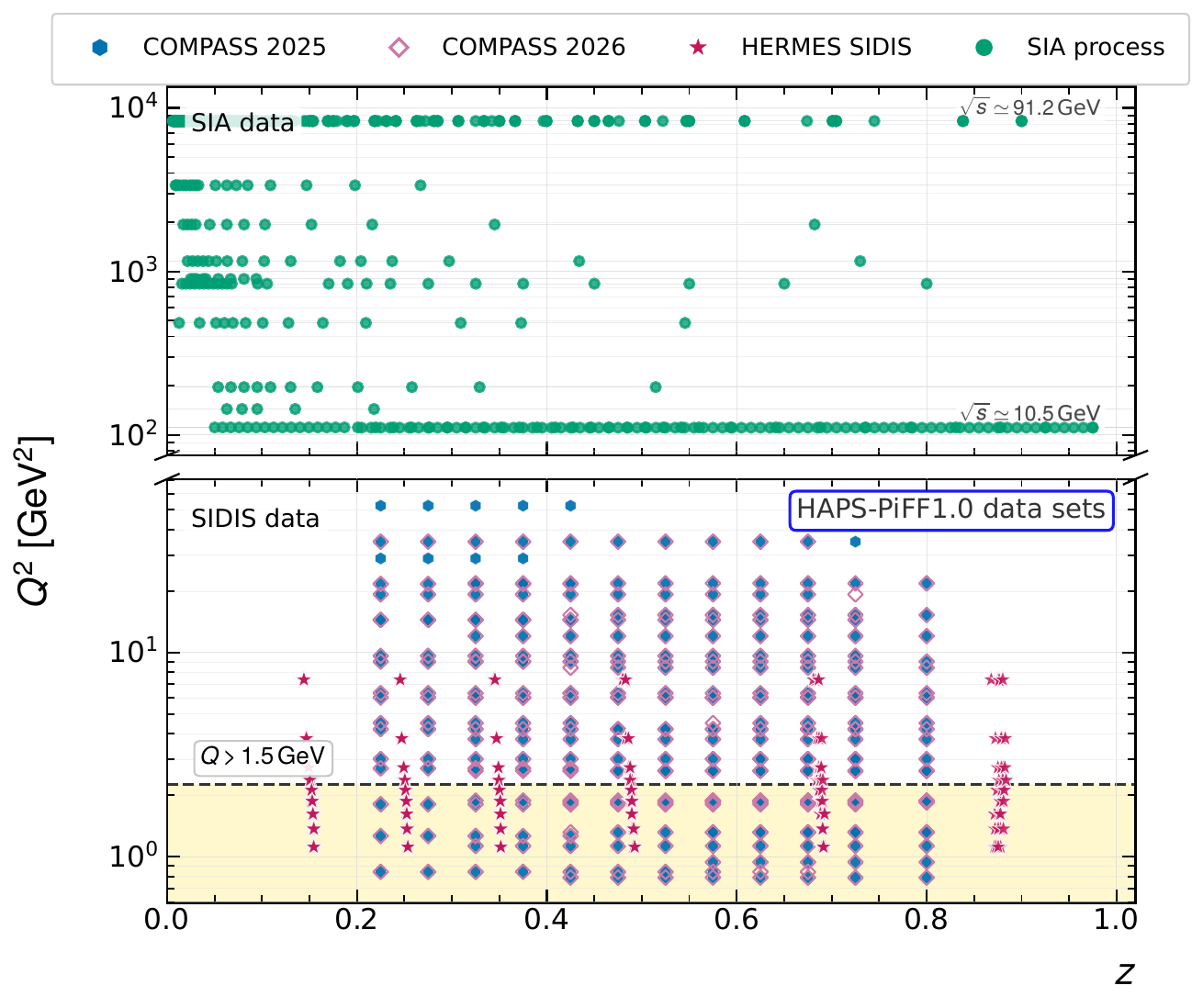} 
\includegraphics[width=0.49\textwidth]{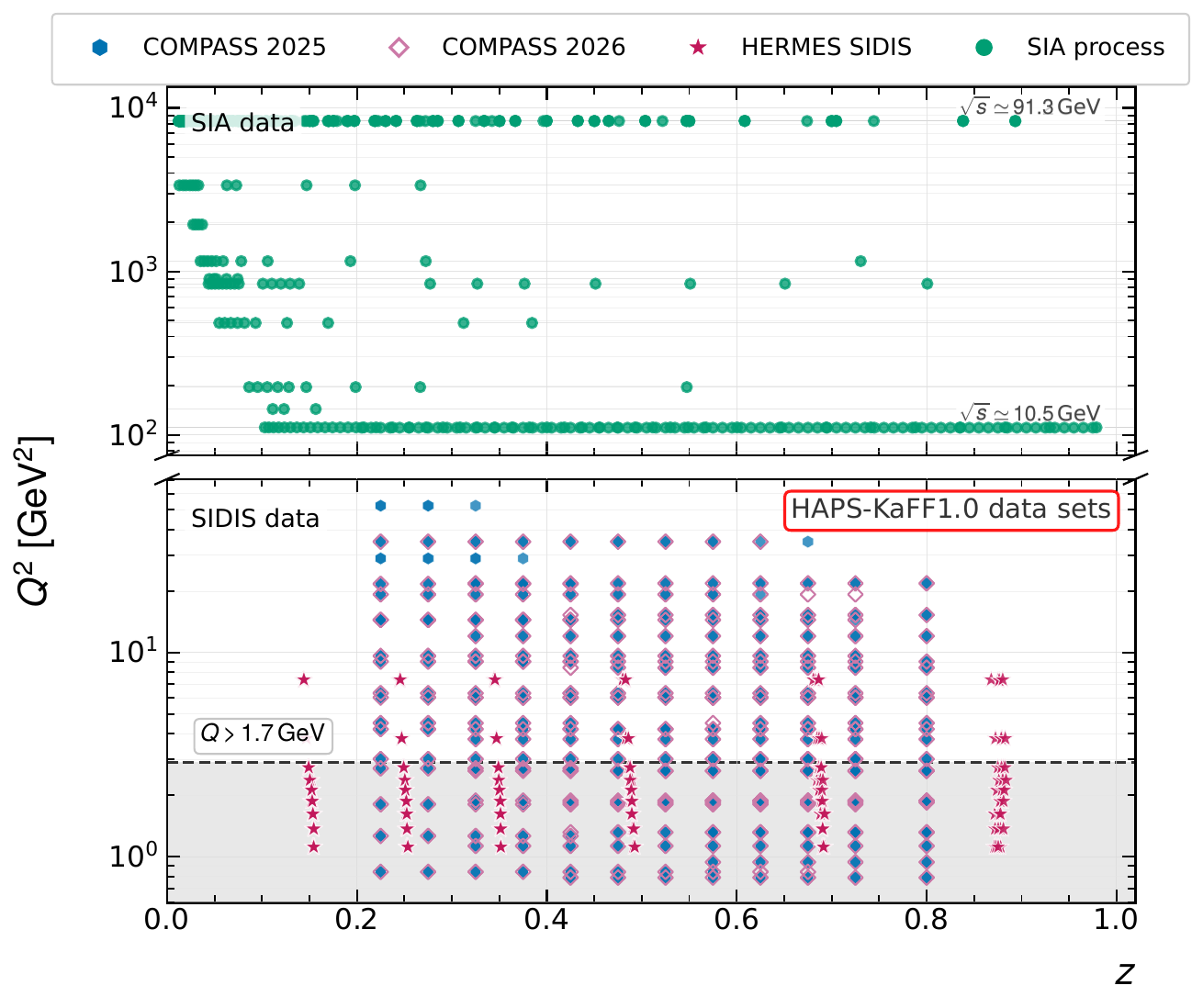} 
\caption{\small \justifying
Kinematic coverage of the charged-pion and charged-kaon datasets
included in the \texttt{HAPS-PiFF1.0} and \texttt{HAPS-KaFF1.0}
analyses in the $(z,Q^2)$ plane. The left and right panels correspond
to the charged-pion and charged-kaon datasets, respectively. In each
panel, the upper part shows the SIA measurements, including low-energy
and $Z$-pole data, while the lower part shows the COMPASS and HERMES SIDIS
multiplicities. The dashed horizontal lines indicate the default lower
SIDIS scale cuts, $Q>1.5~{\rm GeV}$ for the pion analysis and
$Q>1.7~{\rm GeV}$ for the kaon analysis. The broken vertical scale is
used to display the SIA and SIDIS regions in a common figure.}
\label{fig:Q2z}
\end{figure*}

Figure~\ref{fig:Q2z} shows the kinematic coverage of the charged-pion
and charged-kaon data in the $(z,Q^2)$ plane. The left and right panels
correspond to the \texttt{HAPS-PiFF1.0} and \texttt{HAPS-KaFF1.0}
datasets, respectively. In each panel, the upper part displays the SIA
measurements, including low-energy and $Z$-pole data, while the lower
part shows the COMPASS and HERMES SIDIS multiplicities. The figure illustrates the
complementarity between the high-scale SIA data and the lower-scale
SIDIS measurements. The dashed horizontal lines mark the default lower
SIDIS scale cuts used in the final fits, $Q>1.5~{\rm GeV}$ for the pion
analysis and $Q>1.7~{\rm GeV}$ for the kaon analysis. 
These choices are discussed in more detail in Sec.~\ref {Kinematic_cuts}.

\subsection{SIA data sets}

The SIA component of the fit consists of measurements of charged-pion
and charged-kaon production in $e^+e^-$ annihilation. These data are
independent of PDFs and therefore provide direct constraints on
fragmentation functions through charge-summed and electroweak-weighted
combinations. In both analyses, the SIA data are used according to the
normalization, charge definition, and flavor-tagging information
provided by the individual experiments. Depending on the dataset, the
measured observable corresponds to inclusive charged-hadron production,
the charge-summed combinations $\pi^+ + \pi^-$ or $K^+ + K^-$, or a
flavor-tagged sample.

The SIA datasets include measurements from Belle and BaBar at low
center-of-mass energies, together with earlier data from TASSO,
TPC/Two-Gamma, and TOPAZ~\cite{Belle:2013lfg,BaBar:2013yrg,TASSO:1980dyh,TASSO:1982bkc,TASSO:1988jma,TPCTwoGamma:1988yjh,TOPAZ:1994voc,TASSO:1983cre}.
They also include measurements at the $Z$ pole from ALEPH, DELPHI,
OPAL, and SLD~\cite{ALEPH:1994cbg,DELPHI:1998cgx,OPAL:1994zan,SLD:2003ogn}. The
low-energy data constrain the shape of the pion and kaon FFs in the
measured $z$ region and provide scaling-violation information when
combined with higher-scale measurements. The $Z$-pole data provide
high-scale constraints, and the flavor-tagged DELPHI and SLD
measurements help to improve the separation of light- and heavy-quark
fragmentation channels.

After the kinematic selections applied in the present fits, the SIA
component contains $377$ points for the charged-pion analysis and
$339$ points for the charged-kaon analysis. These subsets are displayed
in the upper panel of Fig.~\ref{fig:Q2z}, where the low-energy and
$Z$-pole measurements appear at well-separated values of $Q^2$. Since
SIA data are mostly sensitive to charge-summed and flavor-weighted
combinations, they are not sufficient by themselves to determine the
full flavor decomposition of the charged-pion or charged-kaon FFs. The
SIDIS data described below therefore play an essential complementary
role, especially for separating charge and flavor channels.

The detailed SIA composition is not identical in the pion and kaon
analyses. Some measurements or energy points contribute to one hadron
species but not to the other after the adopted selections, reflecting
the available experimental measurements and the final preprocessing
cuts. The dataset-by-dataset point counts and fit qualities are
therefore reported separately in Table~\ref{tab:haps-kaff-piff-chi2}
and should not be interpreted as implying identical pion and kaon SIA
samples.

\subsection{HERMES and COMPASS SIDIS data}

The SIDIS component of the analysis consists of charge-separated
identified-hadron multiplicities for the processes
$\ell N\to \ell'\pi^\pm X$ and $\ell N\to \ell'K^\pm X$. These
multiplicities are ratios of semi-inclusive to inclusive DIS yields and
therefore depend on both PDFs and FFs. The target dependence and the
comparison of positive and negative hadron production provide flavor
information that is not directly accessible from inclusive SIA data
alone.

The fits include HERMES charged-pion and charged-kaon multiplicities
measured on proton and deuteron targets~\cite{HERMES:2012uyd}. The
HERMES data were collected using a $27.6~{\rm GeV}$ electron/positron
beam incident on hydrogen and deuterium gas targets and were provided
as multidimensional multiplicities in the variables $x_B$, $Q^2$, $z$,
and $P_{h\perp}$. In the present collinear analysis, we use the
charge-separated multiplicity information after applying the projections
and kinematic selections adopted for the fit. The HERMES subset then
contributes $32$ points to the charged-pion fit and $16$ points to the
charged-kaon fit. Although these samples represent a relatively small
fraction of the full datasets, they provide complementary proton and
deuteron constraints on charge-separated identified-hadron production.
Their impact should therefore be interpreted together with the larger
COMPASS SIDIS samples.

The dominant SIDIS input comes from COMPASS. The COMPASS 2025
measurement provides multiplicities for positive and negative pions,
kaons, and unidentified charged hadrons from deep-inelastic scattering
of $160~{\rm GeV}$ muons off a liquid-hydrogen target
~\cite{COMPASS:2024gje}. The data are provided in three-dimensional
bins of $x$, $y$, and $z$ and cover the kinematic region
$Q^2>1~{\rm GeV}^2$, $0.004<x<0.4$, $0.1<y<0.7$, and
$0.2<z<0.85$. The analysis uses improved QED radiative corrections
based on the \textsc{DJANGOH} Monte Carlo generator. In the
\texttt{HAPS-PiFF1.0} fit, the relevant inputs are the charge-separated
proton-target $\pi^+$ and $\pi^-$ multiplicities. After the adopted
kinematic selections, the COMPASS 2025 pion subset contains $222$
$\pi^+$ points and $222$ $\pi^-$ points, corresponding to a total of
$N_{\rm dat}^{\pi}=444$ points. In the \texttt{HAPS-KaFF1.0} fit, the
corresponding inputs are the charge-separated proton-target $K^+$ and
$K^-$ multiplicities. After cuts, the COMPASS 2025 kaon subset contains
$196$ $K^+$ points and $189$ $K^-$ points, giving
$N_{\rm dat}^{K}=385$ points. The proton-target data are especially
useful for constraining flavor combinations weighted by the proton PDFs,
with enhanced sensitivity to $u$-weighted channels. 

The analyses also include the revised COMPASS 2026 isoscalar-target
$\pi^\pm$ and $K^\pm$ multiplicities from the COMPASS addendum~\cite{COMPASS:2025bfn}. 
These data supersede the earlier COMPASS
isoscalar charged-pion, charged-kaon, and unidentified charged-hadron
multiplicities~\cite{COMPASS:2016xvm,COMPASS:2016crr}. The use of the revised
isoscalar datasets is therefore essential for a consistent modern
COMPASS input. After cuts, the COMPASS 2026 pion subset contains $221$
$\pi^+$ points and $221$ $\pi^-$ points, giving
$N_{\rm dat}^{\pi}=442$ points. The COMPASS 2026 kaon subset contains
$197$ $K^+$ points and $197$ $K^-$ points, giving
$N_{\rm dat}^{K}=394$ points.

The proton-target and revised isoscalar-target COMPASS measurements
provide complementary information. The proton data are more directly
sensitive to flavor combinations enhanced by the proton PDFs, while the
isoscalar data provide a more balanced light-flavor weighting. For
charged pions, the comparison of $\pi^+$ and $\pi^-$ multiplicities
mainly constrains favored and unfavored light-quark fragmentation. For
charged kaons, the comparison of $K^+$ and $K^-$ multiplicities adds
sensitivity to strange-to-kaon fragmentation channels. This statement
should not be interpreted as implying that the strange sector is fully
determined by COMPASS alone; correlations with PDFs, unfavored FFs, and
the chosen parametrization remain relevant.

The lower panel of Fig.~\ref{fig:Q2z} displays the COMPASS 2025 and
COMPASS 2026 SIDIS coverage in the $(z,Q^2)$ plane for the charged-pion
and charged-kaon datasets. Compared with the SIA measurements, the SIDIS
data populate lower values of $Q^2$ and provide dense coverage in the
intermediate-$z$ region. The region below the default cut is excluded
from the final fits. This choice is discussed in more detail in the next
subsection.

\subsection{Kinematic cuts}\label{Kinematic_cuts}

Kinematic cuts are applied to define a region in which the fixed-order
leading-twist collinear description is expected to provide a controlled
approximation to the measured observables. For SIA data, cuts on the
hadron energy fraction $z$ are required to avoid regions where the
standard massless collinear description becomes less reliable. At small
$z$, hadron-mass effects and unresummed small-$z$ logarithms may become
important, while at large $z$ the endpoint region can be affected by
threshold logarithms, limited phase space, and possible exclusive or
resonance contributions. For SIDIS data, lower cuts on the hard scale
$Q$ are imposed in order to suppress low-scale regions where higher-twist
effects, low hadronic invariant mass, and the breakdown of the current
fragmentation picture may become relevant. In addition, extreme values
of $z$ can enhance hadron-mass corrections, residual finite-$P_{hT}$
effects in the integrated collinear treatment, target-fragmentation
contributions, and other power-suppressed terms
~\cite{Accardi:2014qda,Guerrero:2015wha,Boglione:2016bph,Boglione:2019nwk}.

\begin{figure*}[t]
\centering
\includegraphics[width=0.48\textwidth]{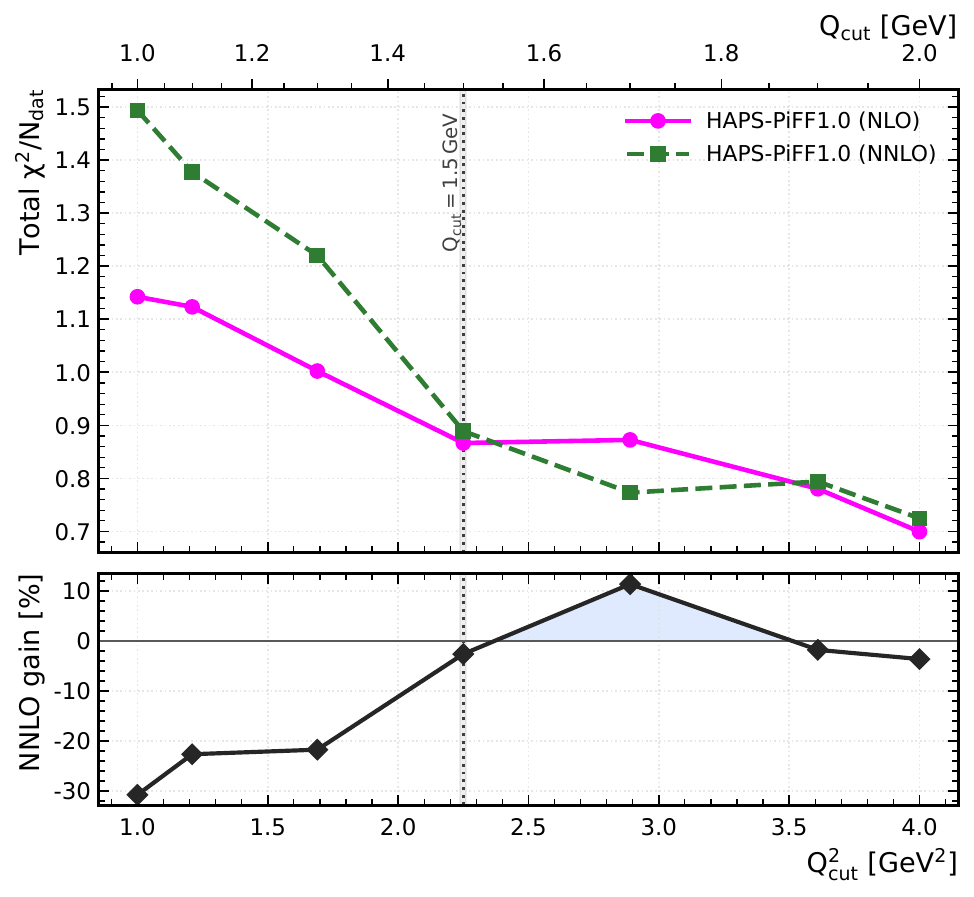}
\hfill
\includegraphics[width=0.48\textwidth]{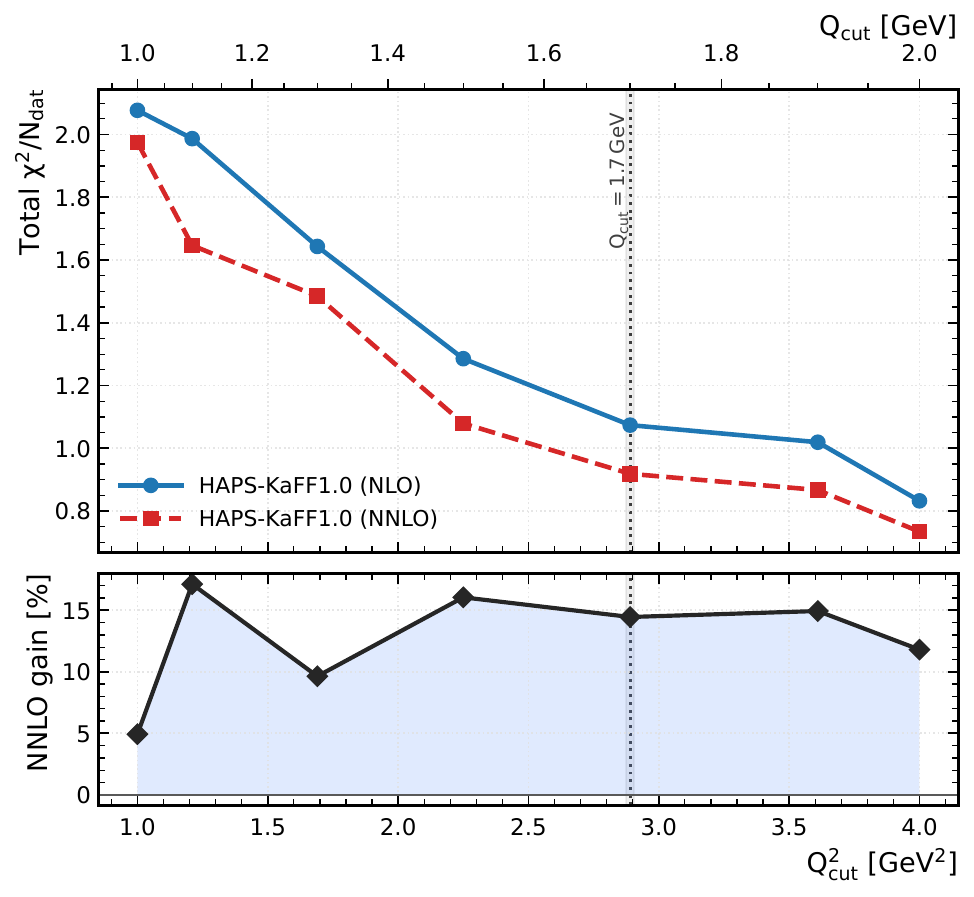}
\caption{\small \justifying
Dependence of the global fit quality on the lower SIDIS scale cut for
the \texttt{HAPS-PiFF1.0} charged-pion fit (left) and the
\texttt{HAPS-KaFF1.0} charged-kaon fit (right). In each case, the upper
subpanel shows $\chi^2/N_{\rm dat}$ at NLO and NNLO as a function of
$Q_{\rm cut}^2$, with the corresponding value of $Q_{\rm cut}$ shown on
the upper axis. The vertical dotted lines indicate the default lower
SIDIS scale cuts used in the final fits, $Q_{\rm cut}=1.5~{\rm GeV}$ for
the pion analysis and $Q_{\rm cut}=1.7~{\rm GeV}$ for the kaon analysis.
The lower subpanels show the relative NNLO gain with respect to the NLO
result; the shaded regions indicate positive NNLO gain, corresponding
to a lower global $\chi^2/N_{\rm dat}$ at NNLO than at NLO.}
\label{fig:Q2scan}
\end{figure*}

In the default fits, lower SIDIS scale cuts are imposed separately for
the charged-kaon and charged-pion analyses. The \texttt{HAPS-KaFF1.0}
fit uses $Q_{\rm cut}^{K}=1.7~{\rm GeV}$, corresponding to
$(Q_{\rm cut}^{K})^2=2.89~{\rm GeV}^2$, while the
\texttt{HAPS-PiFF1.0} fit uses $Q_{\rm cut}^{\pi}=1.5~{\rm GeV}$,
corresponding to $(Q_{\rm cut}^{\pi})^2=2.25~{\rm GeV}^2$. These cuts
are shown by the dashed horizontal line in Fig.~\ref{fig:Q2z} and by
the vertical dotted lines in Fig.~\ref{fig:Q2scan}. 
For the HERMES SIDIS multiplicities, the same lower scale cuts are applied 
together with $0.2<z<0.8$ for both pion and kaon final states. 
The SIA datasets use experiment-dependent cuts in $z$. For the charged-kaon analysis,
the BABAR conventional charged-kaon data are selected with the more
restrictive cut $0.2<z<0.9$. The remaining non-$Z$-pole SIA datasets are
treated with the corresponding experiment-dependent cuts used in the
final runcards, while the $Z$-pole ALEPH, DELPHI, OPAL, and SLD datasets
are retained in the range $0.02<z<0.9$.

The lower SIDIS scale cuts are selected by examining the stability of
the fit quality as a function of $Q^2_{\rm cut}$. Figure~\ref{fig:Q2scan}
shows the global $\chi^2/N_{\rm dat}$ at NLO and NNLO for a range of
lower SIDIS scale cuts. The left panel corresponds to the charged-pion  
analysis, while the right panel corresponds to the charged-kaon analysis.
In each case, the upper subpanel displays the NLO and NNLO profiles as
functions of $Q^2_{\rm cut}$, with the corresponding $Q_{\rm cut}$
values shown on the upper axis. The lower subpanel shows the relative
NNLO gain.
The shaded region highlights the range in which the NNLO gain is positive,
corresponding to a lower global $\chi^2/N_{\rm dat}$ at NNLO than at NLO.

The trends in Fig.~\ref{fig:Q2scan} illustrate the competition between
fit quality and data retention. To make the choice reproducible, for
each hadron species we select the lowest scanned value of $Q_{\rm cut}$
for which the global fit satisfies
$\chi^2_{\rm tot}/N_{\rm dat}\leq 1.1$ at both NLO and NNLO. This
criterion gives $Q_{\rm cut}^{\pi}=1.5~{\rm GeV}$ for pions and
$Q_{\rm cut}^{K}=1.7~{\rm GeV}$ for kaons. It retains the largest
possible SIDIS dataset among the scanned cuts while requiring an
acceptable global description at both perturbative orders.

Increasing the lower scale cut further reduces sensitivity to the
lowest-$Q$ region, where corrections beyond the leading-twist
collinear framework may be more important, but it also removes SIDIS
data that are valuable for charge and flavor separation. For pions,
this information is especially important for constraining favored and
unfavored light-quark fragmentation channels. For kaons, it is
additionally relevant for separating light-quark, unfavored, and
strange-to-kaon fragmentation channels. The selected cuts should not,
however, be interpreted as implying that all nonperturbative effects
are absent above these scales.

\begin{table*}[t]
\centering
\caption{ \small \justifying 
Number of fitted charged-kaon and charged-pion data points,
$N_{\rm dat}^{K}$ and $N_{\rm dat}^{\pi}$, and
$\chi^2/N_{\rm dat}$ values obtained in the
\texttt{HAPS-KaFF1.0} and \texttt{HAPS-PiFF1.0} NLO and NNLO fits.
The table contains the union of the charged-kaon and charged-pion
datasets appearing in the corresponding NLO and NNLO final fit inputs.
The COMPASS 2025 rows correspond to the proton-target multiplicities,
while the COMPASS 2026 rows correspond to the revised isoscalar-target
multiplicities from the COMPASS addendum. A dash indicates that the
dataset is not present for the corresponding hadron species, or that
$N_{\rm dat}=0$ so that $\chi^2/N_{\rm dat}$ is not defined.}
\label{tab:haps-kaff-piff-chi2}
\begin{tabular}{lcrrcccc}
\hline
Experiment & Ref. & $N_{\rm dat}^{K}$ & $N_{\rm dat}^{\pi}$ &
\multicolumn{2}{c}{$K^\pm$} &
\multicolumn{2}{c}{$\pi^\pm$} \\
\cline{5-8}
~&~&~&~& NLO & NNLO & NLO & NNLO \\
\hline
BELLE $h^\pm$                 & \cite{Belle:2013lfg}       & 70 & 70 & 0.39 & 0.40 & 0.14 & 0.12 \\
BABAR conventional $h^\pm$            & \cite{BaBar:2013yrg}       & 28 & -- & 0.66 & 0.25 & -- & -- \\
BABAR prompt $h^\pm$                & \cite{BaBar:2013yrg}       & -- & 39 & -- & -- & 1.58 & 1.31 \\
TASSO 12 GeV $h^\pm$          & \cite{TASSO:1980dyh}       & 3  & 4  & 0.80 & 0.85 & 0.91 & 0.93 \\
TASSO 14 GeV $h^\pm$          & \cite{TASSO:1982bkc}       & 9  & 9  & 1.31 & 1.22 & 1.31 & 1.34 \\
TASSO 22 GeV $h^\pm$          & \cite{TASSO:1982bkc}       & 6  & 8  & 0.75 & 0.89 & 1.63 & 1.79 \\
TPC $h^\pm$                   & \cite{TPCTwoGamma:1988yjh} & 13 & 13 & 0.59 & 0.51 & 0.28 & 0.30 \\
TASSO 30 GeV $h^\pm$          & \cite{TASSO:1983cre}       & 0  & 2  & -- & -- & 0.31 & 0.34 \\
TASSO 34 GeV $h^\pm$          & \cite{TASSO:1988jma}       & 5  & 9  & 0.04 & 0.06 & 1.07 & 1.43 \\
TASSO 44 GeV $h^\pm$          & \cite{TASSO:1988jma}       & 0  & 6  & -- & -- & 1.21 & 1.42 \\
TOPAZ $h^\pm$                 & \cite{TOPAZ:1994voc}       & 3  & 5  & 0.14 & 0.15 & 0.27 & 0.39 \\
ALEPH $h^\pm$                 & \cite{ALEPH:1994cbg}       & 18 & 23 & 0.56 & 0.47 & 1.30 & 1.15 \\
DELPHI inclusive $h^\pm$      & \cite{DELPHI:1998cgx}      & 23 & 21 & 1.04 & 1.08 & 1.17 & 1.25 \\
DELPHI $uds$ tagged $h^\pm$   & \cite{DELPHI:1998cgx}      & 23 & 21 & 0.82 & 0.70 & 2.18 & 2.93 \\
DELPHI $b$ tagged $h^\pm$     & \cite{DELPHI:1998cgx}      & 23 & 21 & 0.63 & 0.48 & 1.88 & 1.78 \\
OPAL $h^\pm$                  & \cite{OPAL:1994zan}        & 10 & 24 & 0.51 & 0.33 & 1.73 & 1.69 \\
SLD inclusive $h^\pm$         & \cite{SLD:2003ogn}         & 35 & 34 & 1.74 & 1.12 & 1.44 & 1.10 \\
SLD $uds$ tagged $h^\pm$      & \cite{SLD:2003ogn}         & 35 & 34 & 1.48 & 1.53 & 1.35 & 2.42 \\
SLD $b$ tagged $h^\pm$        & \cite{SLD:2003ogn}         & 35 & 34 & 1.94 & 1.04 & 0.62 & 0.55 \\
\hline
Total SIA                             &                              & 339 & 377 & 0.96 & 0.75 & 1.11 & 1.19 \\
\hline
HERMES $h^-\, d$                & \cite{HERMES:2012uyd}      & 4   & 8   & 0.56 & 0.29 & 0.30 & 0.04 \\
HERMES $h^-\, p$                & \cite{HERMES:2012uyd}      & 4   & 8   & 0.54 & 0.33 & 0.21 & 0.16 \\
HERMES $h^+\, d$                & \cite{HERMES:2012uyd}      & 4   & 8   & 1.15 & 0.65 & 0.29 & 0.07 \\
HERMES $h^+\, p$                & \cite{HERMES:2012uyd}      & 4   & 8   & 1.78 & 0.97 & 0.21 & 0.35 \\
\hline
Total HERMES                          &                              & 16  & 32  & 1.01 & 0.56 & 0.25 & 0.16 \\
\hline
COMPASS 2026 $h^+$              & \cite{COMPASS:2025bfn}     & 197 & 221 & 1.23 & 0.94 & 1.12 & 1.11 \\
COMPASS 2026 $h^-$              & \cite{COMPASS:2025bfn}     & 197 & 221 & 0.99 & 0.86 & 0.76 & 0.90 \\
\hline
Total COMPASS 2026                    &                              & 394 & 442 & 1.11 & 0.90 & 0.94 & 1.01 \\
\hline
COMPASS 2025 $h^+$              & \cite{COMPASS:2024gje}     & 196 & 222 & 1.65 & 1.19 & 0.92 & 0.75 \\
COMPASS 2025 $h^-$              & \cite{COMPASS:2024gje}     & 189 & 222 & 1.17 & 1.14 & 0.74 & 0.74 \\
\hline
Total COMPASS 2025                    &                              & 385 & 444 & 1.41 & 1.17 & 0.83 & 0.74 \\
\hline
Total SIDIS                           &                              & 795 & 918 & 1.25 & 1.02 & 0.86 & 0.85 \\
\hline
\textbf{Global data set}              &                              & \textbf{1134} & \textbf{1295} &
\textbf{1.13} & \textbf{0.89} & \textbf{0.8723} & \textbf{0.8742} \\
\hline
\end{tabular}
\end{table*}

\section{Results}\label{sec:results}

In this section we present the main results of the
\texttt{HAPS-PiFF1.0} and \texttt{HAPS-KaFF1.0} analyses. The discussion
is organized around three related questions: how well the combined
SIA+SIDIS datasets are described at NLO and NNLO, how the modern COMPASS
SIDIS input affects the charged-pion and charged-kaon determinations, and
how stable the extracted flavor-dependent FFs are under the inclusion of
NNLO corrections. We first discuss the global and dataset-level fit
quality, then examine the description of the COMPASS 2025 proton-target
and COMPASS 2026 revised isoscalar-target multiplicities. We finally
compare the resulting NLO and NNLO fragmentation functions and discuss
their relation to existing charged-pion and charged-kaon FF
determinations.

\subsection{Fit quality}

Table~\ref{tab:haps-kaff-piff-chi2} summarizes the dataset-level fit
quality for the \texttt{HAPS-KaFF1.0} and \texttt{HAPS-PiFF1.0}
analyses at NLO and NNLO. The table reports the number of fitted
charged-kaon and charged-pion data points, $N_{\rm dat}^{K}$ and
$N_{\rm dat}^{\pi}$, together with the corresponding
$\chi^2/N_{\rm dat}$ values for each perturbative order. The rows
include the SIA and SIDIS datasets entering either the kaon or pion fit.
A dash indicates that the corresponding dataset is absent for that
hadron species, or that no fitted points remain after the final
kinematic selections. The COMPASS 2025 rows correspond to the
proton-target multiplicities, while the COMPASS 2026 rows correspond to
the revised isoscalar-target multiplicities from the COMPASS addendum.

For the charged-kaon analysis, the NNLO fit provides a clear improvement
relative to NLO. The global $\chi^2/N_{\rm dat}$ decreases from $1.13$
at NLO to $0.89$ at NNLO. This reduction is visible in both the SIA and
SIDIS sectors. The SIA contribution decreases from $0.96$ to $0.75$,
while the total SIDIS contribution decreases from $1.25$ to $1.02$.
The dominant COMPASS subsets also improve: the COMPASS 2025
proton-target contribution decreases from $1.41$ to $1.17$, and the
COMPASS 2026 revised isoscalar-target contribution decreases from
$1.11$ to $0.90$. Among the COMPASS kaon subsets, the largest
improvement is observed for the COMPASS 2025 $K^+$ multiplicities, for
which $\chi^2/N_{\rm dat}$ decreases from $1.65$ to $1.19$.

The charged-pion analysis displays a different NLO-NNLO pattern. The
global $\chi^2/N_{\rm dat}$ remains essentially unchanged, with values
of $0.8723$ at NLO and $0.8742$  at NNLO. The total SIDIS
contribution is also stable, changing only from $0.86$ to $0.85$. The
COMPASS 2025 proton-target pion contribution improves from $0.83$ to
$0.74$, while the COMPASS 2026 revised isoscalar-target contribution
increases from $0.94$ to $1.01$. The HERMES pion subset improves from
$0.25$ to $0.16$, although its statistical weight in the global fit is
limited compared with the COMPASS samples. In contrast to the kaon case,
the pion SIA contribution increases from $1.11$ to $1.19$. The pion
results therefore should not be interpreted as a uniform improvement at
NNLO. Instead, they indicate a redistribution of fit quality among
experimental sectors, with improvements in some SIDIS subsets
compensated by mild deteriorations in SIA and in the COMPASS 2026
revised isoscalar subset.

\begin{figure*}[t]
\centering
\includegraphics[width=0.48\textwidth]{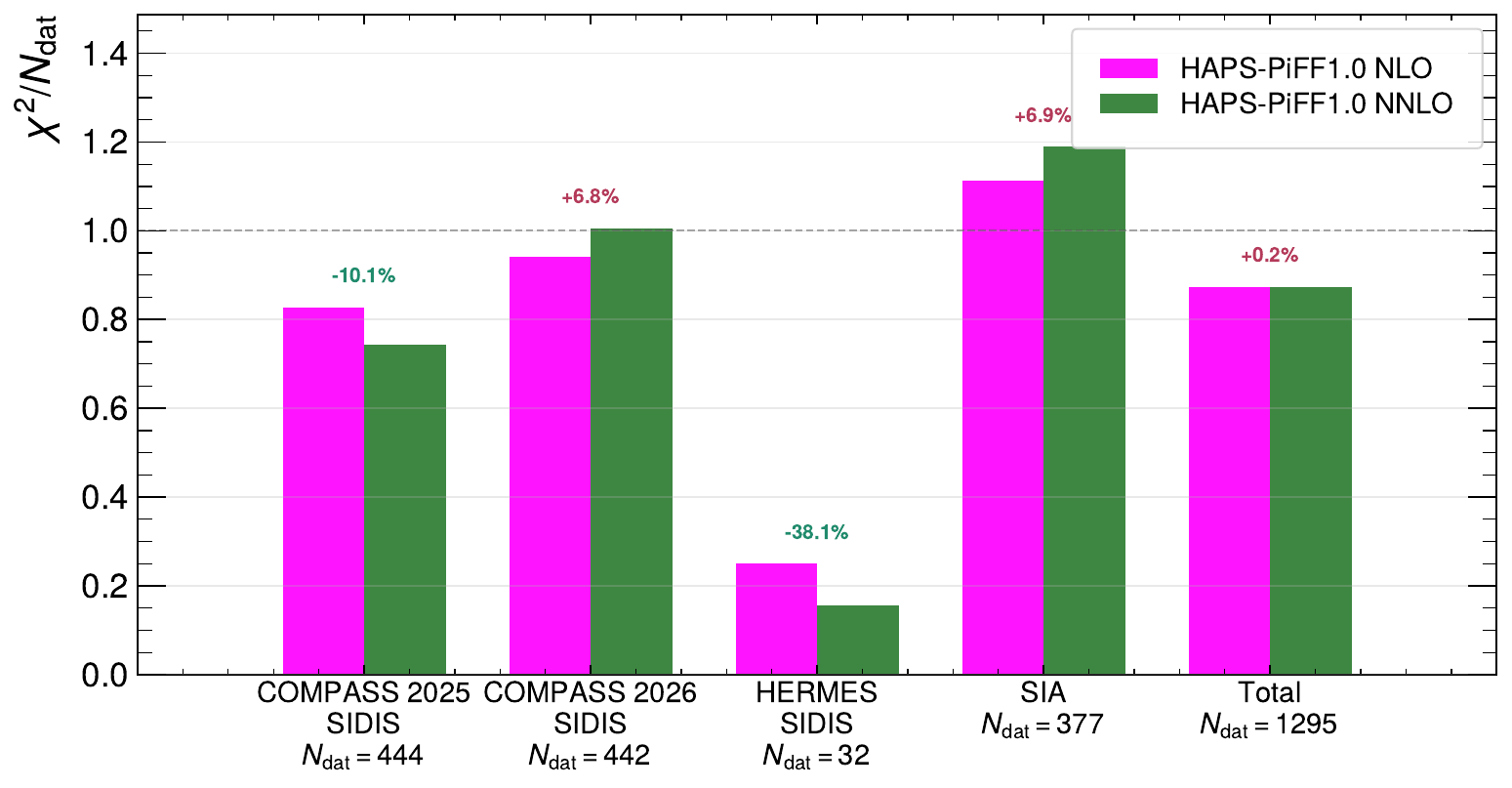}
\hfill
\includegraphics[width=0.48\textwidth]{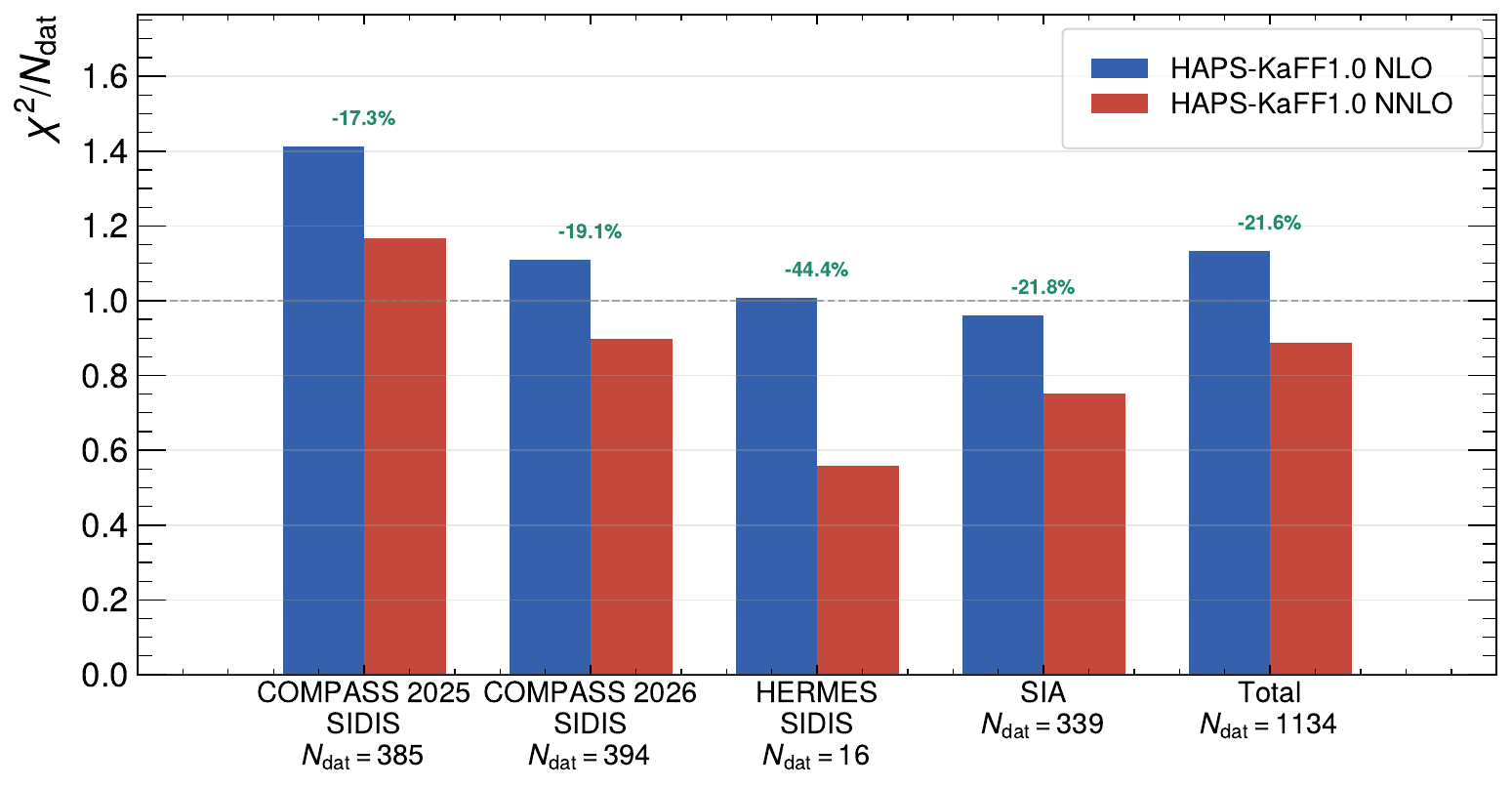}
\caption{\small \justifying
Grouped comparison of the fit quality obtained in the
\texttt{HAPS-PiFF1.0} charged-pion  analysis (left) and the
\texttt{HAPS-KaFF1.0} charged-kaon analysis (right) at NLO and NNLO.
The bars show $\chi^2/N_{\rm dat}$ for the COMPASS 2025 proton-target
SIDIS data, the COMPASS 2026 revised isoscalar-target SIDIS data,
HERMES SIDIS data, SIA data, and the global dataset. The percentage
labels indicate the relative NNLO--NLO change.}
\label{fig:chi2_NLO_NNLO_1}
\end{figure*}

The grouped behavior is shown in Fig.~\ref{fig:chi2_NLO_NNLO_1}. The
left panel corresponds to the \texttt{HAPS-PiFF1.0} charged-pion 
analysis, while the right panel corresponds to the \texttt{HAPS-KaFF1.0}
charged-kaon analysis. The percentage labels quantify the relative
NNLO--NLO change in each group.

For kaons, the group-level comparison confirms the improvement already
seen in Table~\ref{tab:haps-kaff-piff-chi2}: all displayed kaon groups
move toward smaller $\chi^2/N_{\rm dat}$ values at NNLO. For pions, the
pattern is more mixed. The COMPASS 2025 and HERMES groups improve, the
global fit quality remains nearly unchanged, and the COMPASS 2026 and
SIA groups show mild increases. This behavior reinforces the conclusion
that the NNLO pion fit involves compensating shifts among data sectors
rather than a uniform improvement of all fitted subsets.

\begin{figure*}[t]
\centering
\includegraphics[width=0.48\textwidth]{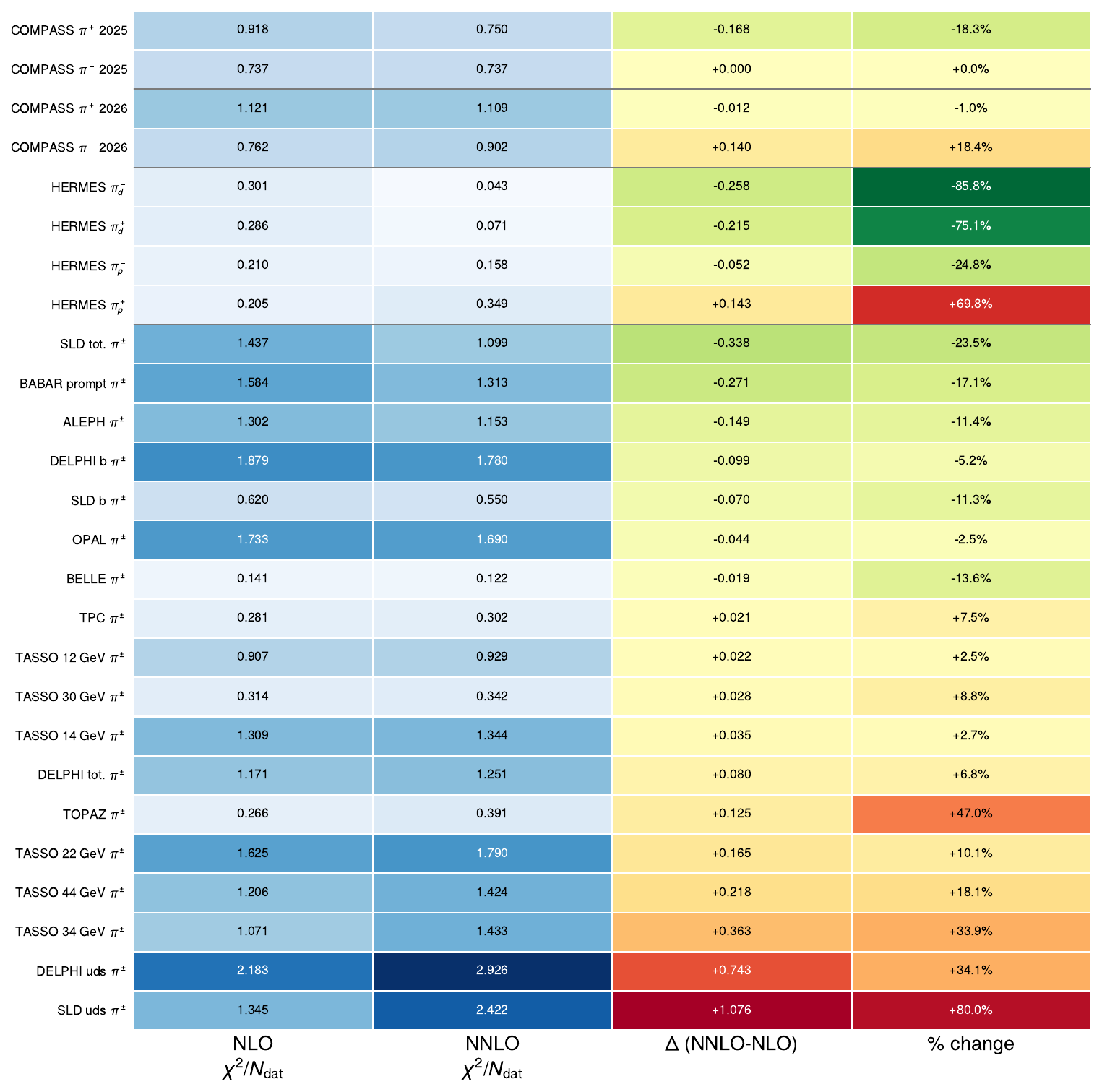}
\hfill
\includegraphics[width=0.48\textwidth]{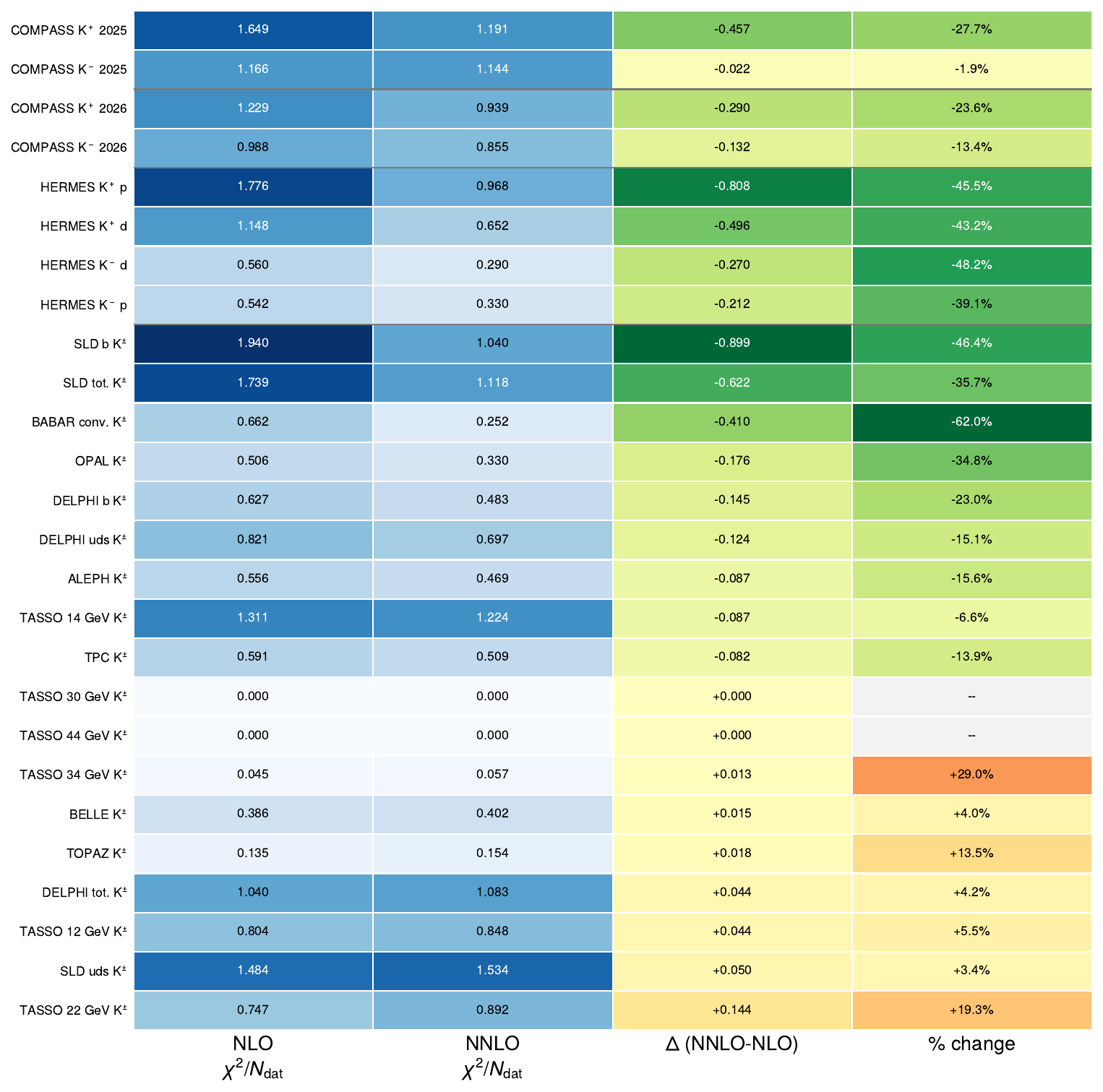}
\caption{\small \justifying
Dataset-level comparison of $\chi^2/N_{\rm dat}$ at NLO and NNLO for
the \texttt{HAPS-PiFF1.0} charged-pion analysis (left) and the
\texttt{HAPS-KaFF1.0} charged-kaon analysis (right). In each matrix, the
first two columns show the NLO and NNLO values, while the remaining
columns show the absolute and relative NNLO--NLO changes. Negative
relative changes correspond to an improved description at NNLO.}
\label{fig:chi2_NLO_NNLO_2}
\end{figure*}

The dataset-level matrices in Fig.~\ref{fig:chi2_NLO_NNLO_2} provide a
more differential view of these trends. For the charged-kaon analysis,
the matrix highlights sizeable NNLO improvements for BABAR, SLD
inclusive, SLD $b$-tagged, the HERMES subsets, and the COMPASS $K^+$
multiplicities, while a small number of datasets mildly deteriorate at
NNLO. For the charged-pion analysis, the corresponding matrix shows that
the nearly unchanged global $\chi^2/N_{\rm dat}$ results from
nonuniform shifts among individual SIA and SIDIS subsets. This
dataset-level information is important because the global
$\chi^2/N_{\rm dat}$ alone does not reveal how the fit quality is
redistributed across experiments.

We next examine the description of the COMPASS multiplicities directly.

\subsection{Description of COMPASS multiplicities}\label{subsec:compass-multiplicities}

\begin{figure*}[t]
\centering
\includegraphics[width=0.80\textwidth]{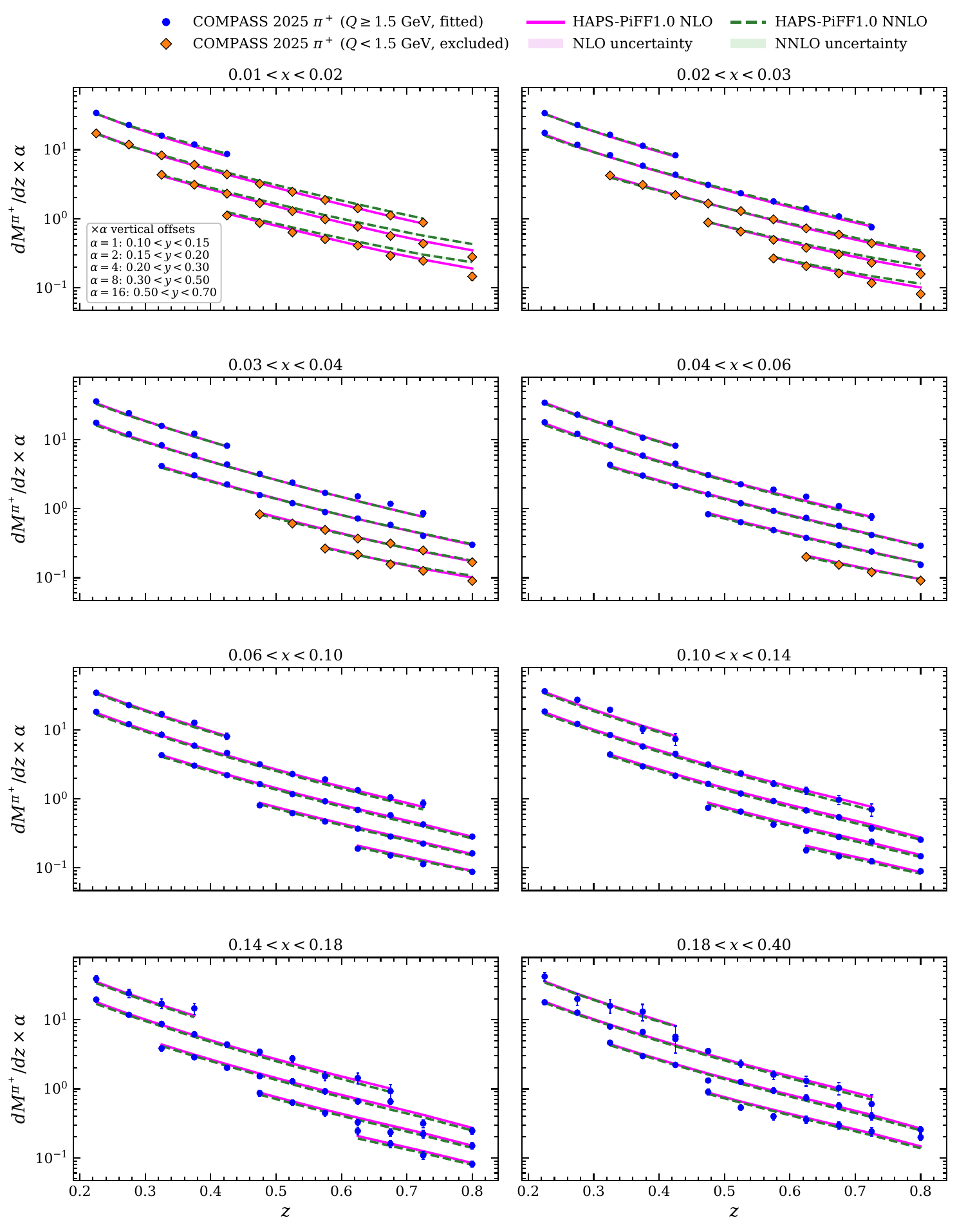}
\caption{\small \justifying
Comparison of the \texttt{HAPS-PiFF1.0} NLO and NNLO predictions with
the COMPASS 2025 proton-target $\pi^+$ multiplicities~\cite{COMPASS:2024gje}.
The data are shown as functions of $z$ in bins of $x$ and $y$. For each
$y$ bin, the experimental multiplicities, the NLO and NNLO predictions,
and their uncertainty bands are multiplied by the same factor $\alpha$
specified in the inset legend. This rescaling is introduced only to
separate the different $y$-bin series vertically. The shaded bands represent the one-standard-deviation FF-replica uncertainty
and are often barely visible because of their small width.}
\label{fig:compass2025-pion}
\end{figure*}

\begin{figure*}[t]
\centering
\includegraphics[width=0.80\textwidth]{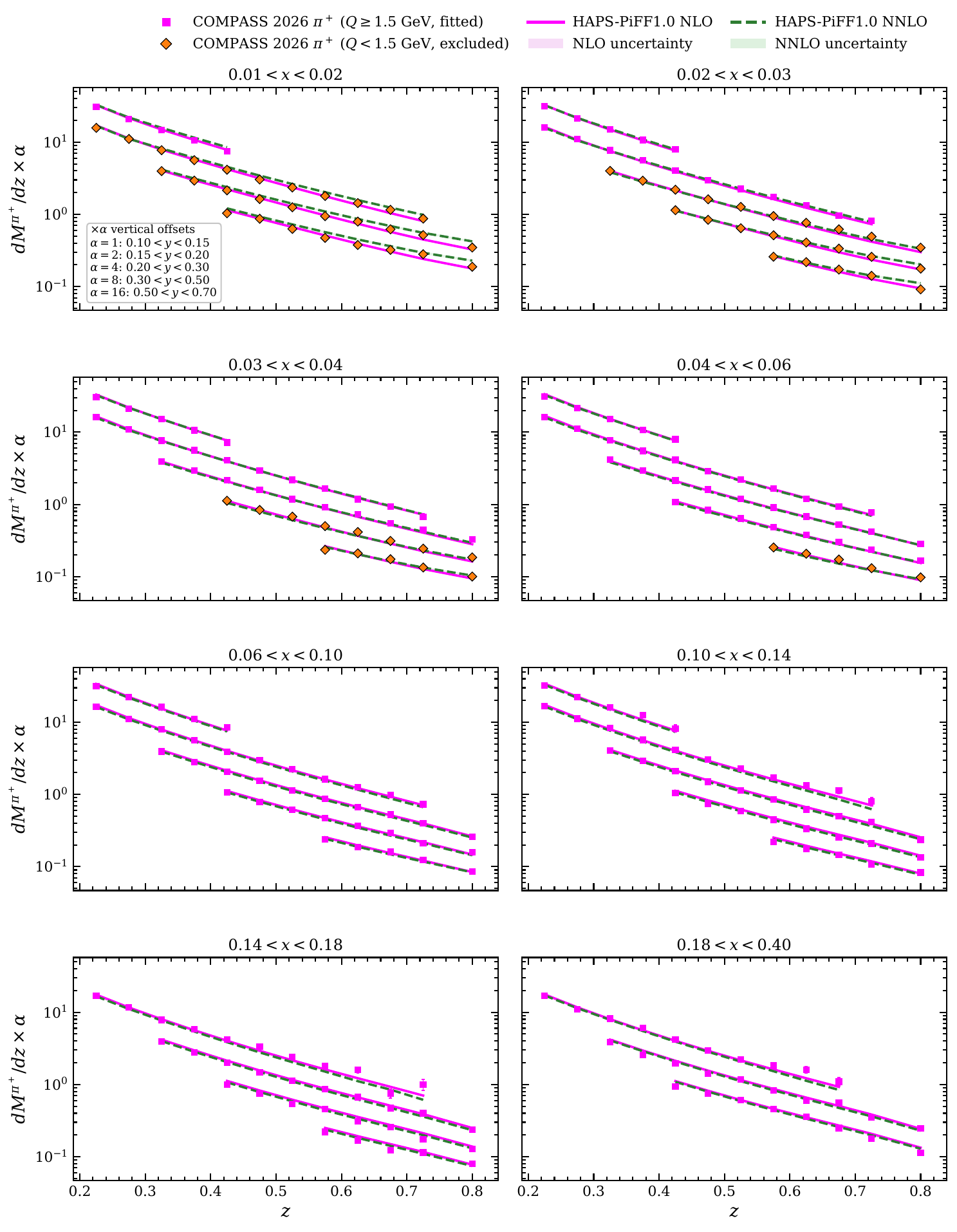}
\caption{\small \justifying
Same as Fig.~\ref{fig:compass2025-pion}, but for the COMPASS 2026
revised isoscalar-target $\pi^+$ multiplicities from the COMPASS
addendum~\cite{COMPASS:2025bfn}. }
\label{fig:compass2026-pion}
\end{figure*}

\begin{figure*}[t]
\centering
\includegraphics[width=0.80\textwidth]{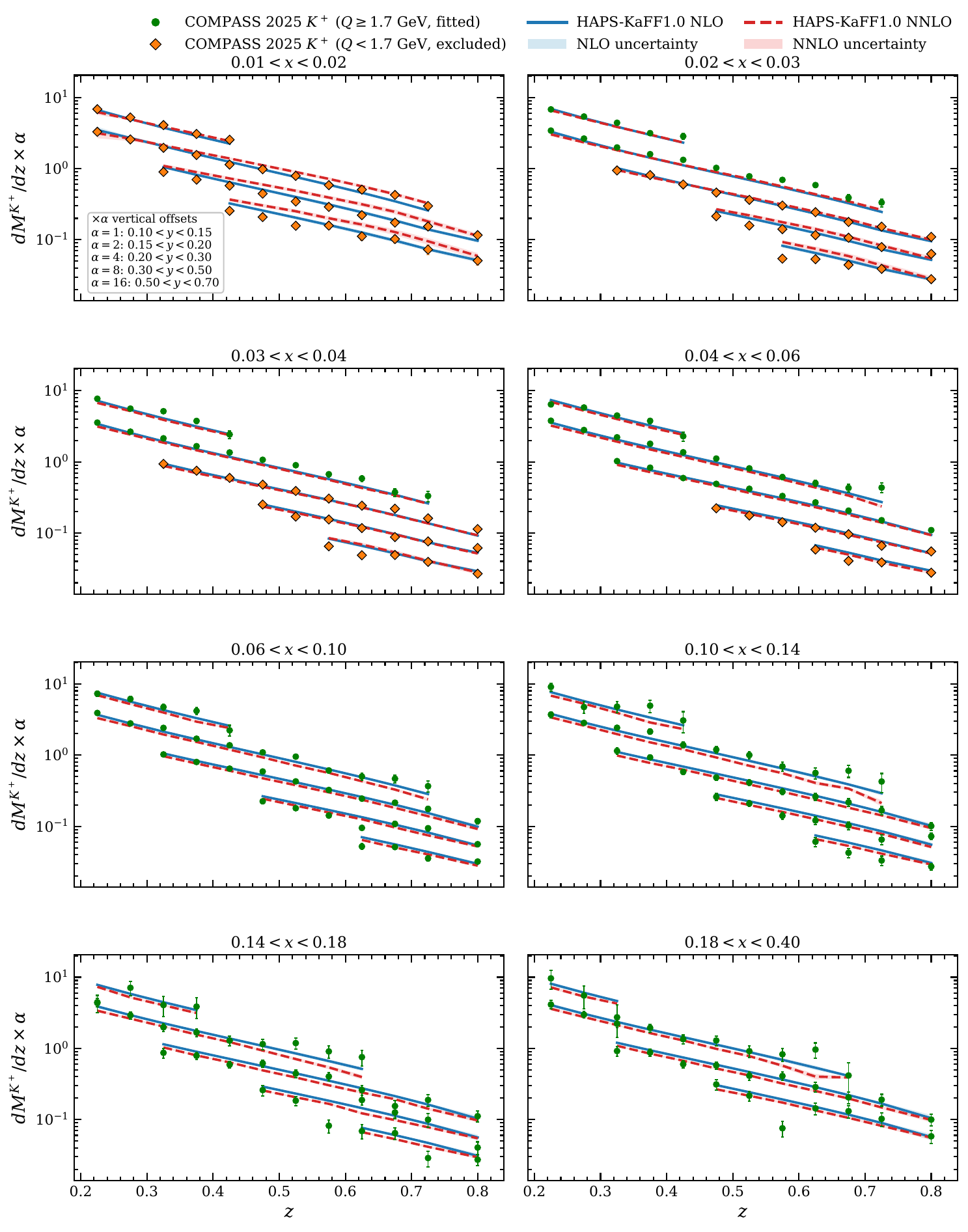}
\caption{\small \justifying
Comparison of the \texttt{HAPS-KaFF1.0} NLO and NNLO predictions with
the COMPASS 2025 proton-target $K^+$ multiplicities~\cite{COMPASS:2024gje}.
The data are shown as functions of $z$ in bins of $x$ and $y$. 
All data points in the displayed $x$ range are shown; points with $Q<1.7$~GeV, excluded from the fit, are indicated separately.
For each $y$ bin, the experimental multiplicities, the NLO and NNLO predictions,
and their uncertainty bands are multiplied by the same factor $\alpha$
specified in the inset legend. This rescaling is introduced only to
separate the different $y$-bin series vertically. The shaded bands represent the one-standard-deviation FF-replica uncertainty
and are often barely visible because of their small width.}

\label{fig:compass2025-kaon}
\end{figure*}

\begin{figure*}[t]
\centering
\includegraphics[width=0.80\textwidth]{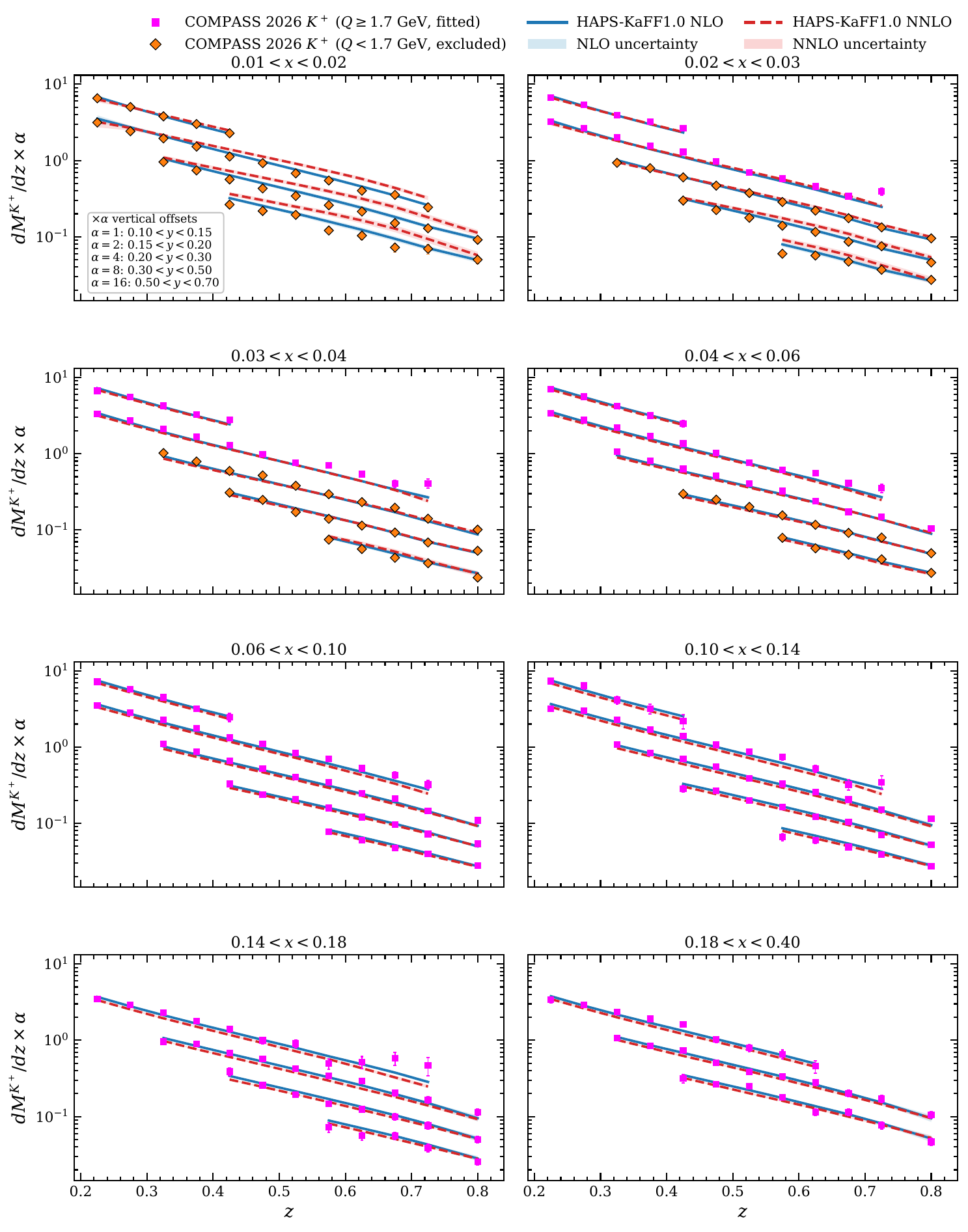}
\caption{\small \justifying
Same as Fig.~\ref{fig:compass2025-kaon}, but for the COMPASS 2026
revised isoscalar-target $K^+$ multiplicities from the COMPASS
addendum~\cite{COMPASS:2025bfn}. }
\label{fig:compass2026-kaon}
\end{figure*}

We now examine the description of the modern COMPASS multiplicities more
directly. Figures~\ref{fig:compass2025-pion}--\ref{fig:compass2026-kaon}
show representative comparisons between the fitted NLO and NNLO
predictions and the COMPASS 2025 proton-target and COMPASS 2026 revised
isoscalar-target multiplicities. The observables are displayed as
functions of $z$ in bins of $x$ and $y$, following the experimental
binning. All data
points in the displayed kinematic range are shown, with points included in
the fit clearly distinguished from those excluded by the corresponding
kinematic cuts. 
The COMPASS 2025 proton-target data provide flavor sensitivity
weighted by the proton PDFs, while the COMPASS 2026 revised isoscalar
multiplicities provide a more balanced light-flavor weighting and
supersede the earlier COMPASS isoscalar measurements.

For visual clarity, the different $y$-bin series within each fixed-$x$
panel are separated vertically: for a given $y$ bin, the measured
multiplicities, the corresponding NLO and NNLO predictions, and their
uncertainty bands are all multiplied by the same factor $\alpha$
specified in the inset legend. Thus, $\alpha$ is purely a plotting
rescaling and does not modify the relative data-theory comparison.

For charged pions, the COMPASS multiplicities mainly constrain favored
and unfavored light-quark fragmentation. The COMPASS 2025 $\pi^+$
comparison in Fig.~\ref{fig:compass2025-pion} is consistent with the
improvement of the corresponding table entry from
$\chi^2/N_{\rm dat}=0.92$ at NLO to $0.75$ at NNLO. In contrast, the
COMPASS 2026 revised isoscalar $\pi^+$ contribution, shown in
Fig.~\ref{fig:compass2026-pion}, remains essentially stable, changing
only from $1.12$ to $1.11$. This behavior is consistent with the broader
pattern discussed above: the pion fit remains globally stable from NLO
to NNLO, with improvements in some subsets compensated by mild
deteriorations or near-stability in others.

For charged kaons, the COMPASS multiplicities are particularly relevant
for the separation of favored and unfavored light-quark fragmentation
and for constraining the strange-to-kaon channel. The comparison with
the COMPASS 2025 $K^+$ multiplicities in
Fig.~\ref{fig:compass2025-kaon} reflects the improvement already seen at
the level of the fit quality in Table~\ref{tab:haps-kaff-piff-chi2},
where the corresponding contribution decreases from
$\chi^2/N_{\rm dat}=1.65$ at NLO to $1.19$ at NNLO. A similar pattern is
observed for the COMPASS 2026 revised isoscalar $K^+$ multiplicities,
shown in Fig.~\ref{fig:compass2026-kaon}, for which the corresponding
entry decreases from $1.23$ to $0.94$. These improvements indicate that
the NNLO kaon fit provides a more consistent description of the positive
kaon multiplicities in both proton and revised isoscalar targets.

We note that in some of the lowest-$x$ panels of
Figs.~\ref{fig:compass2025-kaon} and \ref{fig:compass2026-kaon},
the NNLO curves may appear visually less close to the data than the NLO curves.
However, many of the points in these low-$x$ panels, and in particular most
of those responsible for this visual impression, lie below the kaon cut
$Q=1.7$~GeV and are therefore excluded from the fit. These points are now
clearly distinguished in the figures from the data entering the fit and
should not be used to assess the fitted $\chi^2$. For the fitted data,
Table~\ref{tab:haps-kaff-piff-chi2} shows that the COMPASS 2025 and
COMPASS 2026 $K^+$ contributions to $\chi^2/N_{\rm dat}$ both improve
at NNLO, together with the total kaon SIDIS and global fit quality.

In all panels, the shaded bands represent the uncertainty propagated
from the fitted FF replica ensemble. They should not be interpreted as a
complete theory uncertainty, since they do not include variations of the
factorization scale, alternative PDF inputs, hadron-mass effects, or
other possible power-suppressed contributions. The comparison therefore
provides a direct diagnostic of the fitted replica uncertainty and the
relative NLO--NNLO stability within the adopted fitting framework.

\clearpage

\subsection{Sensitivity to the reduced kaon flavor basis}\label{subsec:kaon-basis-stability}

To assess the dependence of the strange-sector determination on the
flavor assumptions adopted in the baseline kaon fit, we performed an
additional NLO fit in which the constraint
\[
D_s^{K^+}=D_{\bar u}^{K^+}
\]
is relaxed and the two FFs are parametrized independently. The second
relation,
\[
D_d^{K^+}=D_{\bar d}^{K^+},
\]
is kept unchanged. All other ingredients of the analysis, including the
experimental datasets, kinematic cuts, PDF input, perturbative setup, and
training strategy, are kept identical to those of the baseline
\texttt{HAPS-KaFF1.0} NLO determination. The purpose of this variant fit
is therefore to isolate the sensitivity of the extracted strange-sector
FFs to this particular reduced-basis assumption.

The additional flavor freedom leads to a moderate improvement in the
overall fit quality. For the same $N_{\rm dat}=1134$ data points, the
global value changes from
\[
\frac{\chi^2}{N_{\rm dat}}=1.131
\]
in the baseline NLO fit to
\[
\frac{\chi^2}{N_{\rm dat}}=1.071
\]
when $D_s^{K^+}$ and $D_{\bar u}^{K^+}$ are fitted independently. 
Since the relaxed fit contains additional free parameters, this
improvement should not by itself be interpreted as evidence against the
baseline flavor constraint. Rather, the relevant question is whether the
extracted strange-to-kaon FFs remain stable under the enlarged flavor
basis.

\begin{figure*}[t]
\centering
\includegraphics[width=0.48\textwidth]
{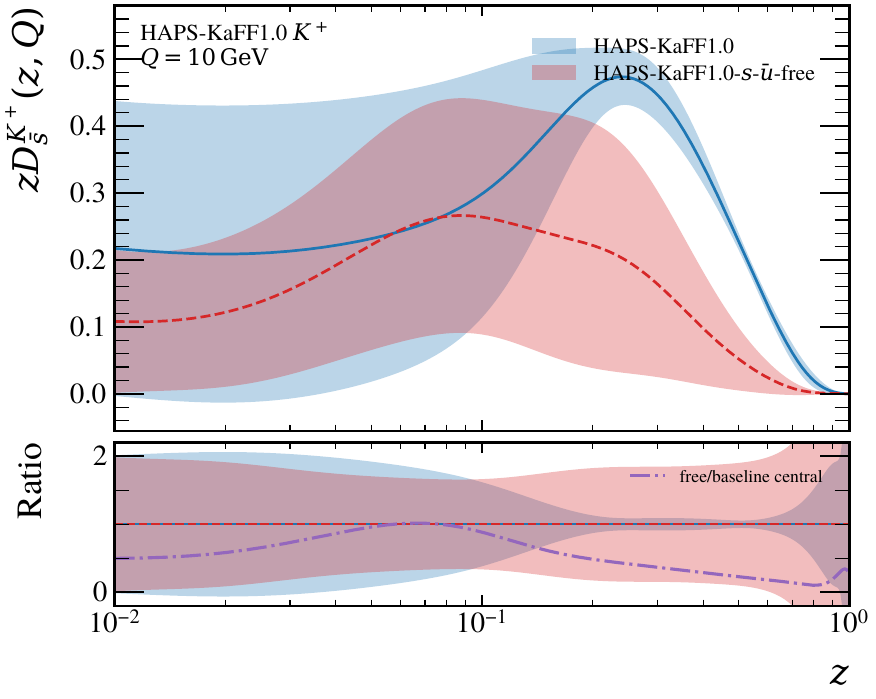}
\hfill
\includegraphics[width=0.48\textwidth]
{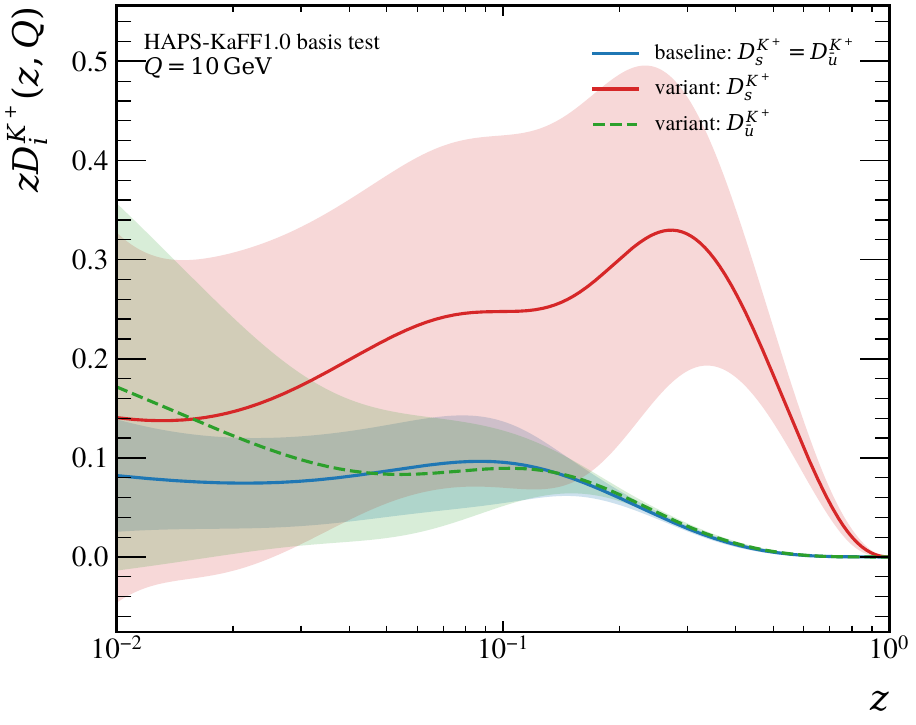}
\caption{\small \justifying  
Sensitivity of the charged-kaon FFs to the reduced flavor-basis
assumption at $Q=10~\mathrm{GeV}$. Left: comparison of
$zD_{\bar s}^{K^+}(z,Q)$ obtained in the baseline
\texttt{HAPS-KaFF1.0} NLO fit and in the variant fit in which
$D_s^{K^+}$ and $D_{\bar u}^{K^+}$ are parametrized independently.
The bands denote the corresponding one-standard-deviation Monte Carlo
uncertainties. Right: comparison of the common
$D_s^{K^+}=D_{\bar u}^{K^+}$ distribution imposed in the baseline fit
with the independently fitted $D_s^{K^+}$ and $D_{\bar u}^{K^+}$
distributions in the relaxed-basis fit.} 
\label{fig:kaon-basis-stability}
\end{figure*}

The comparison is shown in Fig.~\ref{fig:kaon-basis-stability}. The
right panel demonstrates that, once the equality
$D_s^{K^+}=D_{\bar u}^{K^+}$ is removed, the additional flexibility is
used predominantly by the strange channel. In particular,
$D_s^{K^+}$ develops a substantially larger central value over the
intermediate-$z$ region, whereas $D_{\bar u}^{K^+}$ remains much closer
to the common distribution obtained in the baseline fit. This behavior
shows that the two channels are not equally constrained by the present
SIA and SIDIS dataset once they are allowed to vary independently.

The consequence for the favored strange-to-kaon channel is shown in the
left panel. Although $D_{\bar s}^{K^+}$ is already parametrized
independently in the baseline analysis, relaxing the
$D_s^{K^+}=D_{\bar u}^{K^+}$ relation induces a visible shift in its
central value and a substantial broadening of its uncertainty over part
of the phenomenologically relevant $z$ region. The effect reflects the
correlations among the strange and light-antiquark channels entering the
charge-separated kaon multiplicities. By contrast, the independently
fitted $D_{\bar u}^{K^+}$ distribution remains broadly compatible with
the corresponding baseline result.

This stability study therefore indicates that the COMPASS and HERMES
SIDIS measurements provide important sensitivity to the strange sector,
but that the quantitative extraction of $D_{\bar s}^{K^+}$ retains a
non-negligible dependence on the adopted flavor basis. Accordingly, the
strange-to-kaon constraint obtained in the baseline
\texttt{HAPS-KaFF1.0} analysis should be interpreted within the stated
flavor assumptions rather than as completely parametrization
independent.

\subsection{NLO versus NNLO fragmentation functions}\label{subsec:nlo-nnlo-ffs}

We now compare the NLO and NNLO fragmentation functions obtained in the
\texttt{HAPS-PiFF1.0} and \texttt{HAPS-KaFF1.0} analyses. This
comparison provides a direct diagnostic of perturbative stability in the
extracted FFs. Since the NLO and NNLO fits are performed independently,
differences between the two sets reflect the combined effect of the
higher-order coefficient functions, timelike evolution, refitting of the
nonperturbative parametrization, and the constraints imposed by the
fitted SIA and SIDIS datasets.

\begin{figure*}[t]
\centering
\includegraphics[width=0.30\textwidth]{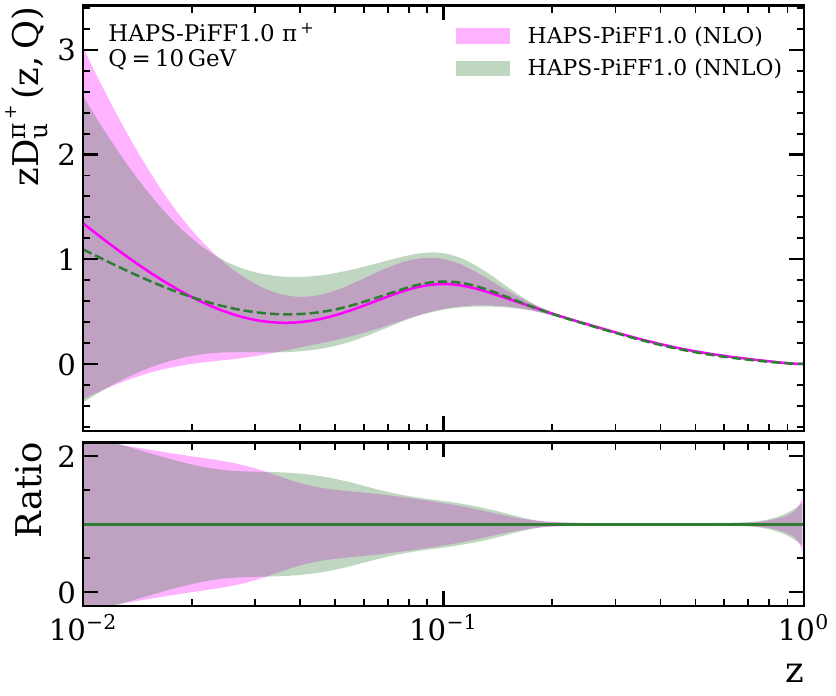}
\includegraphics[width=0.30\textwidth]{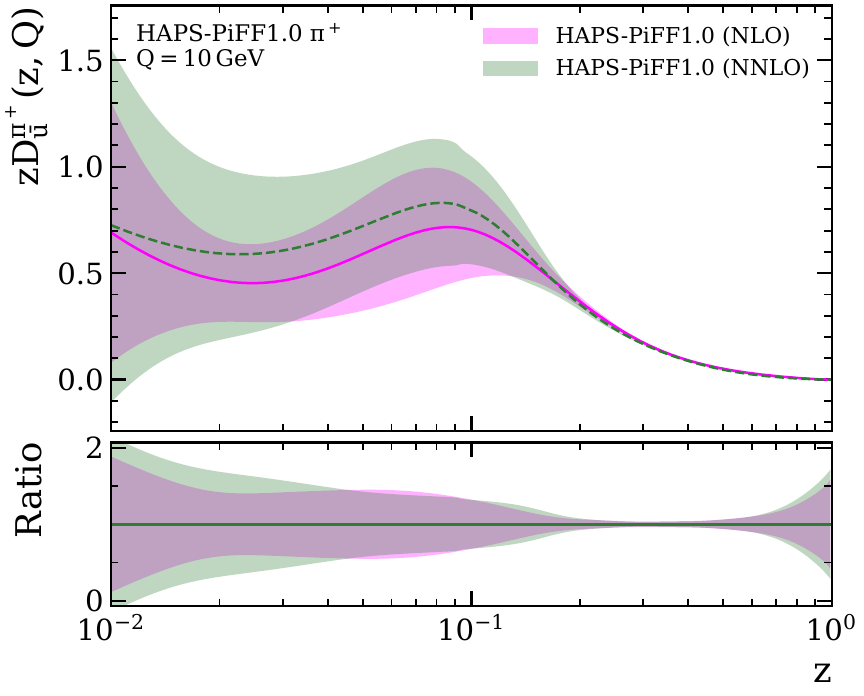}
\includegraphics[width=0.30\textwidth]{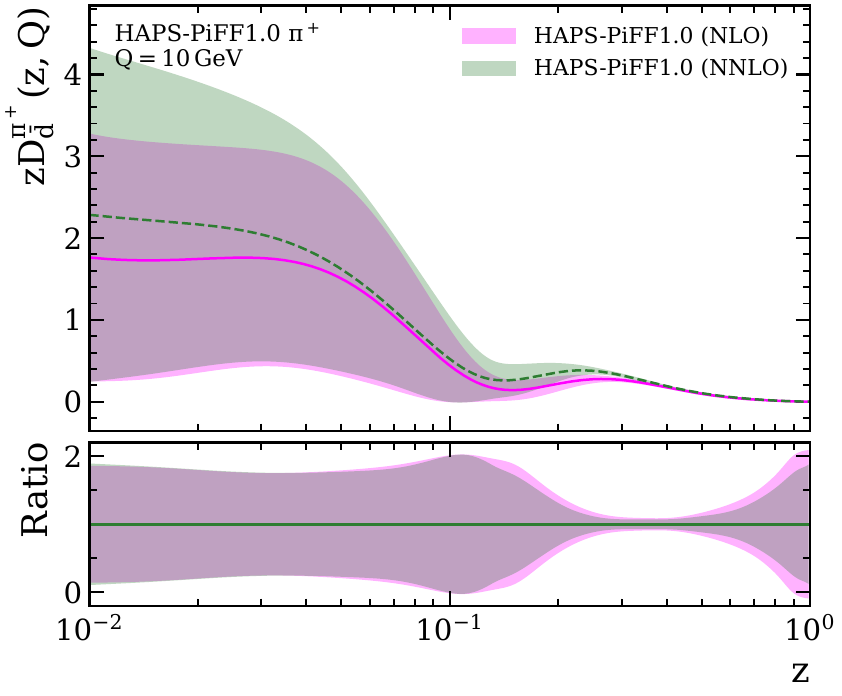}
\vspace{0.2cm}
\includegraphics[width=0.30\textwidth]{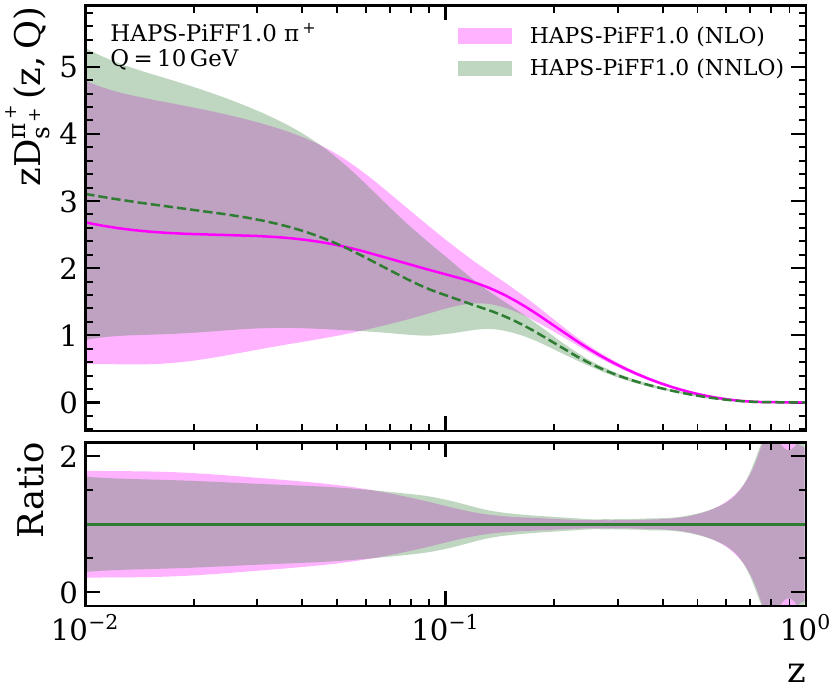}
\includegraphics[width=0.30\textwidth]{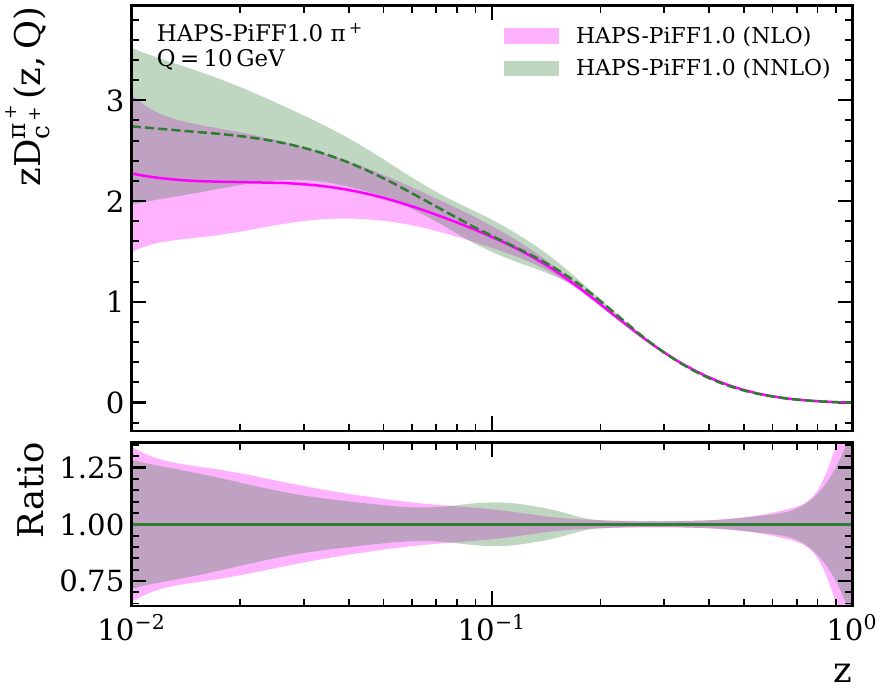}
\includegraphics[width=0.30\textwidth]{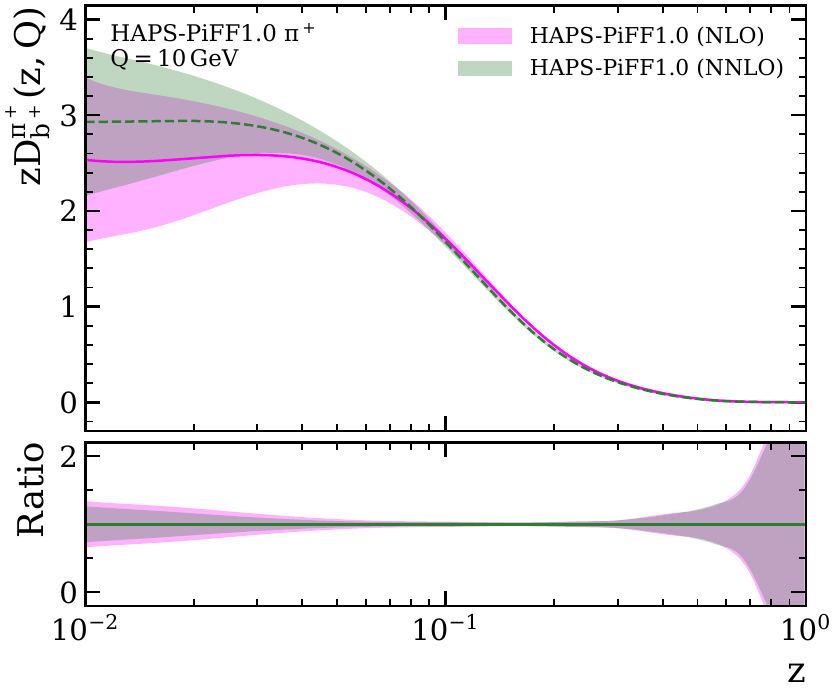}
\vspace{0.2cm}
\includegraphics[width=0.30\textwidth]{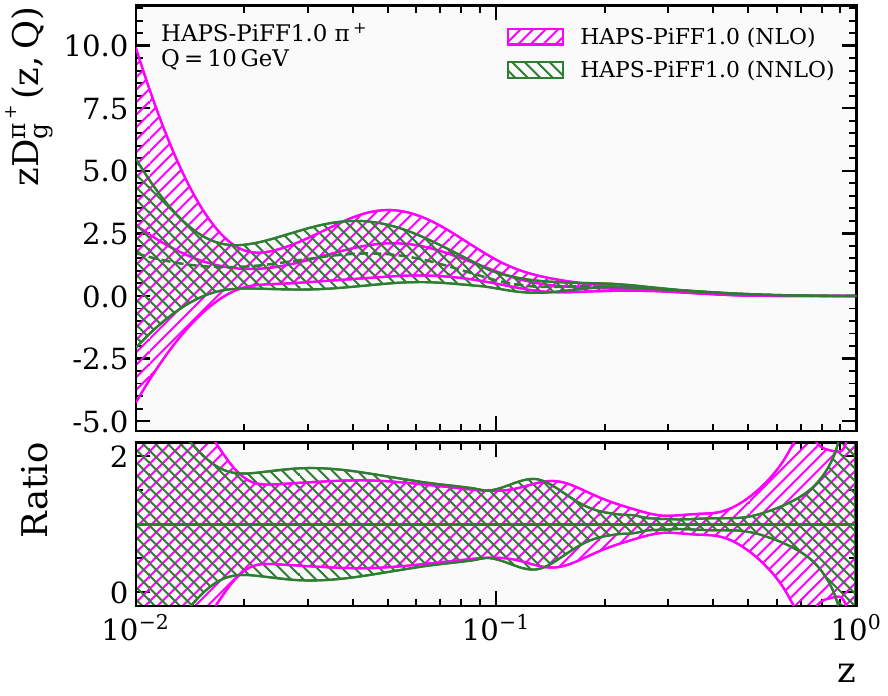}
\caption{\small \justifying
Comparison of the \texttt{HAPS-PiFF1.0} charged-pion fragmentation
functions at NLO and NNLO for $\pi^+$ production at $Q=10~{\rm GeV}$.
The panels show $zD_i^{\pi^+}(z,Q)$ for the indicated parton flavors:
$u$, $\bar u$, $\bar d$, $s^+$, $c^+$, $b^+$, and $g$. The bands denote
the one-standard-deviation Monte Carlo replica uncertainties. 
The lower ratio panel in each plot shows the NLO and NNLO uncertainty
bands normalized to their respective central values.
The gluon panel is hatched to emphasize that the gluon FF is only
indirectly constrained in the present SIA+SIDIS fit. }
\label{fig:piff-nlo-nnlo}
\end{figure*}

Figure~\ref{fig:piff-nlo-nnlo} shows the charged-pion FFs for
$\pi^+$ production at $Q=10~{\rm GeV}$. The displayed combinations
include light-quark, strange, heavy-quark, and gluon FFs. The pion
sector is primarily constrained by the interplay of SIA data with
charge-separated SIDIS multiplicities, which provide sensitivity to
favored and unfavored light-quark fragmentation. The comparison indicates
that the NLO and NNLO pion FFs are broadly compatible within the fitted
replica uncertainties in the data-constrained region. 
Larger relative uncertainty bands, where present, should be interpreted
with care, especially in regions of $z$ where the experimental
constraints are weaker or where the corresponding central FF becomes
small. In such regions, the self-normalized ratio panels can visually
enhance the apparent size of the replica uncertainty.

\begin{figure*}[t]
\centering
\includegraphics[width=0.30\textwidth]{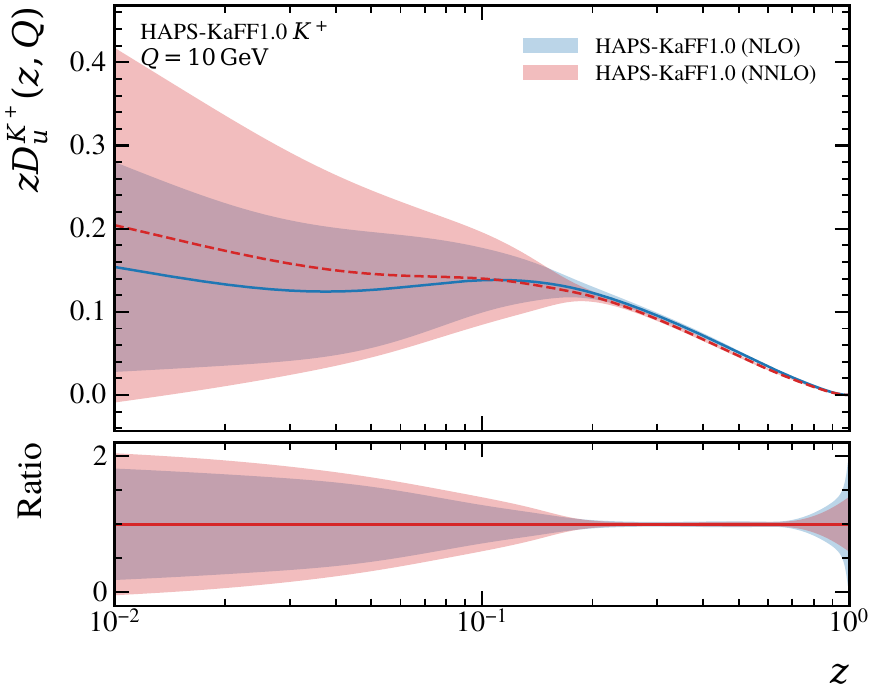}
\includegraphics[width=0.30\textwidth]{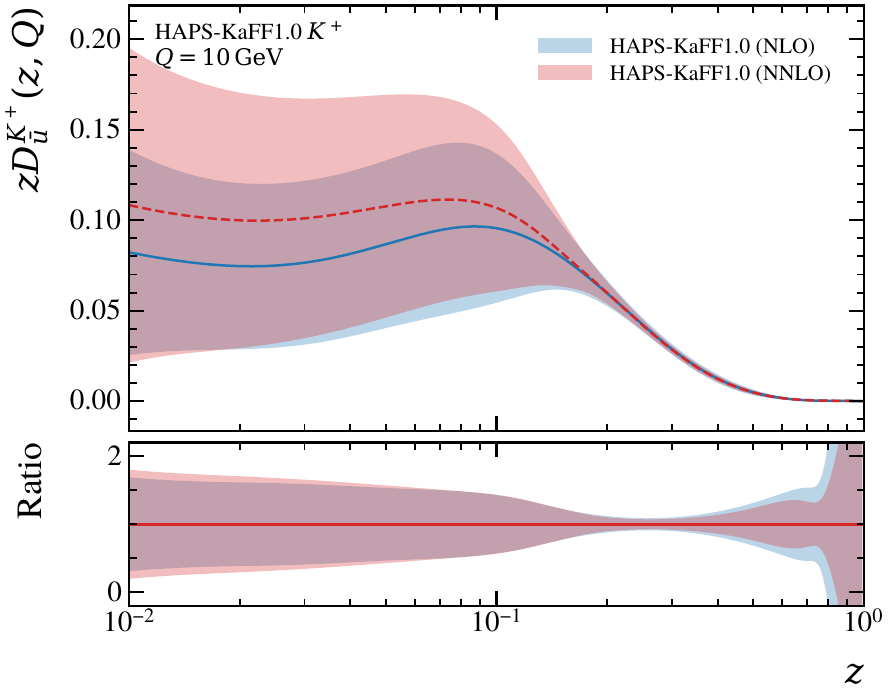}
\includegraphics[width=0.30\textwidth]{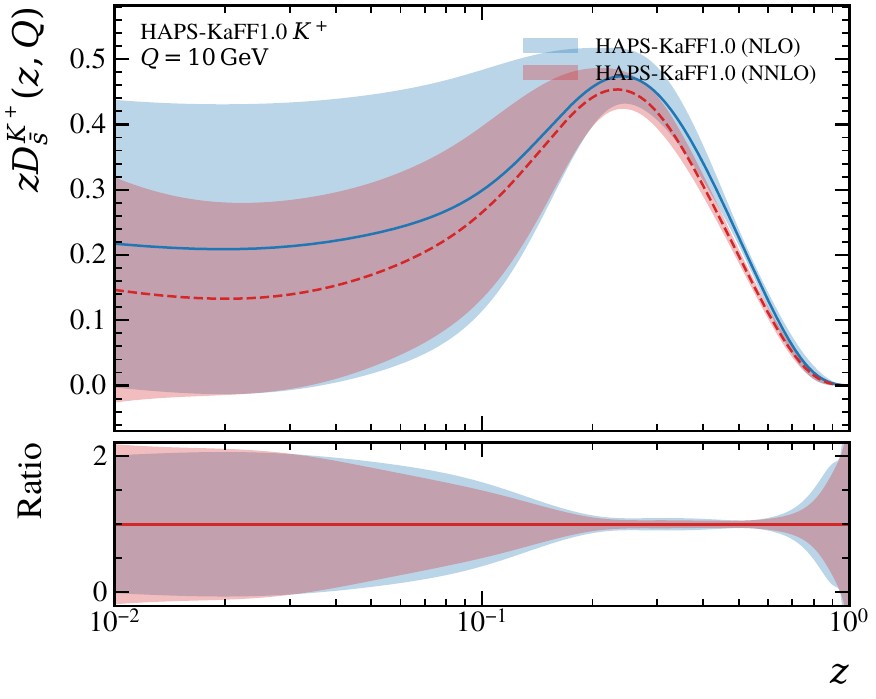}
\vspace{0.2cm}
\includegraphics[width=0.30\textwidth]{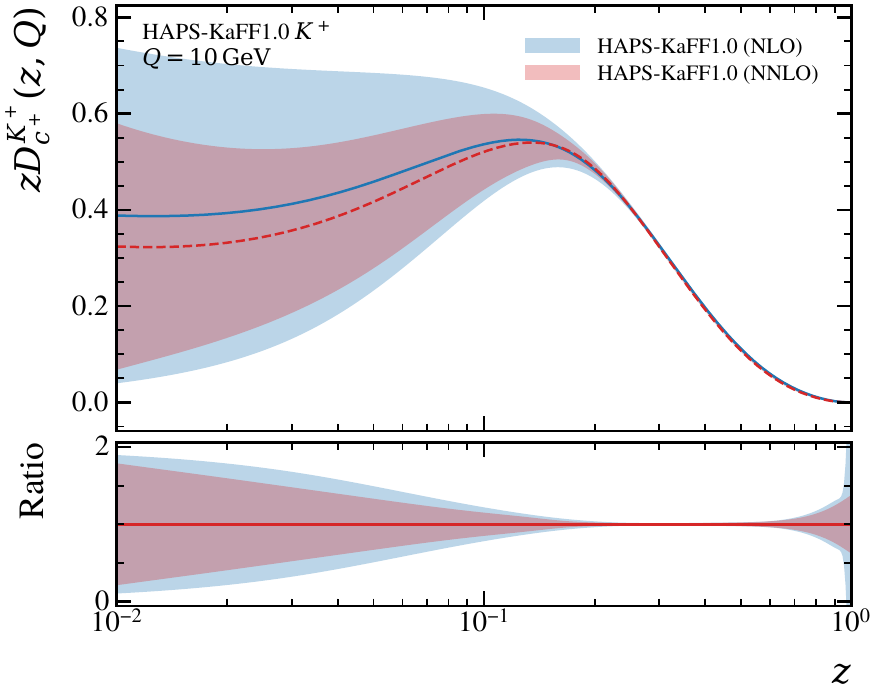}
\includegraphics[width=0.30\textwidth]{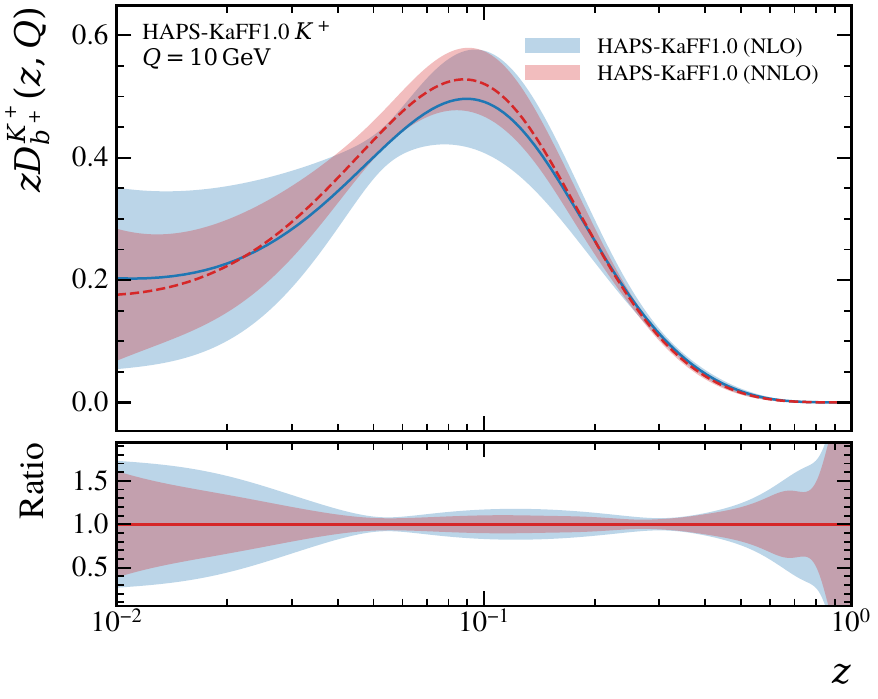}
\includegraphics[width=0.30\textwidth]{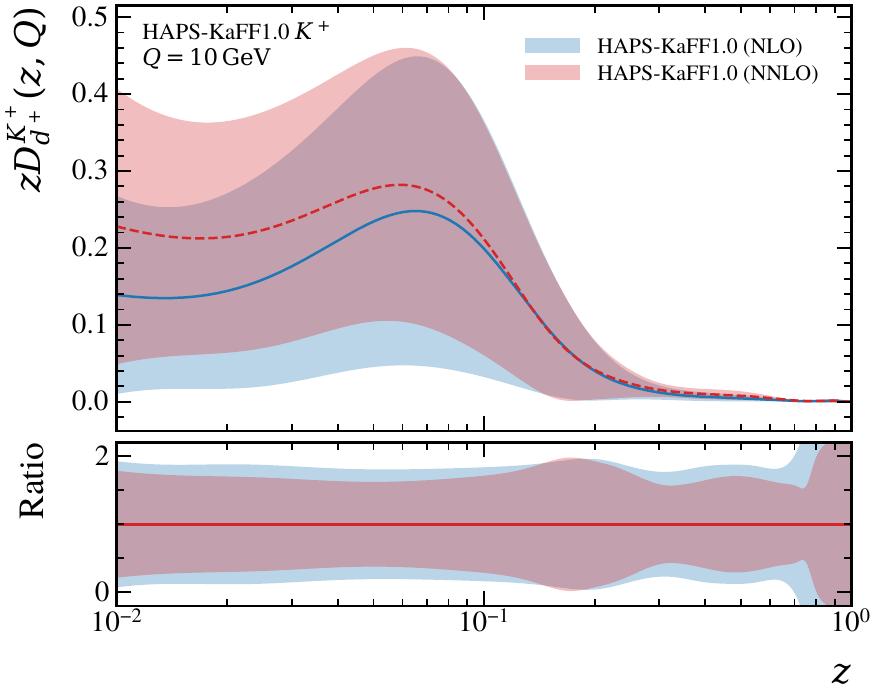}
\vspace{0.2cm}
\includegraphics[width=0.30\textwidth]{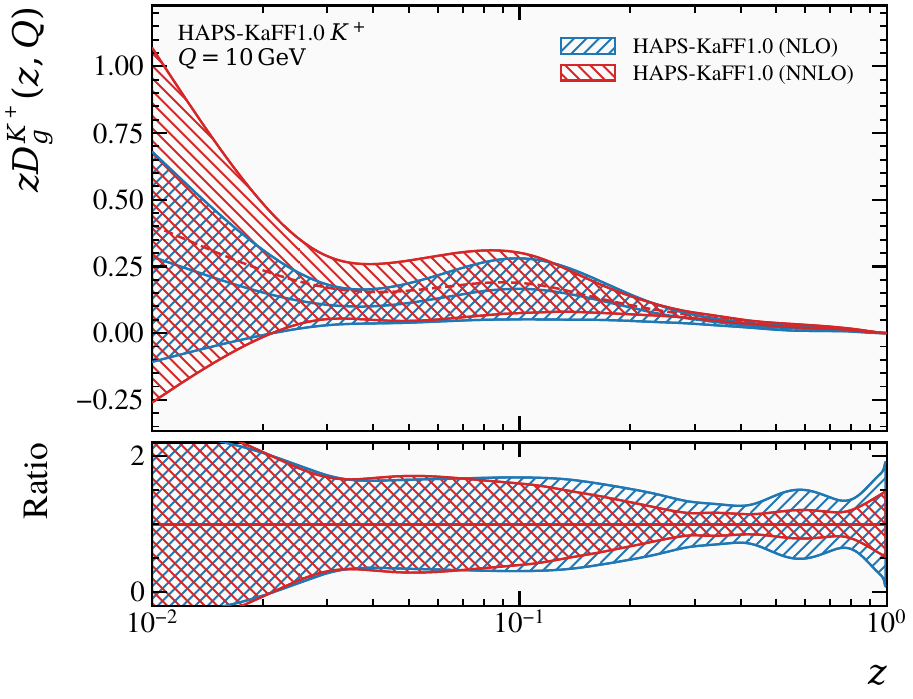} 
\caption{\small \justifying
Comparison of the \texttt{HAPS-KaFF1.0} charged-kaon fragmentation
functions at NLO and NNLO for $K^+$ production at $Q=10~{\rm GeV}$.
The panels show $zD_i^{K^+}(z,Q)$ for the indicated parton flavors:
$u$, $\bar u$, $\bar s$, $c^+$, $b^+$, $d^+$, and $g$. The bands denote
the one-standard-deviation Monte Carlo replica uncertainties. 
The lower ratio panels show the NLO and NNLO uncertainty bands normalized
to their corresponding central values.
The gluon panel is hatched to indicate its only indirect constraint in
the present SIA+SIDIS fit. }
\label{fig:kaff-nlo-nnlo}
\end{figure*}

The charged-kaon FFs are shown in Fig.~\ref{fig:kaff-nlo-nnlo}. In this
case, the phenomenologically important channels include the favored
$u\to K^+$ and $\bar{s}\to K^+$ fragmentation functions, together with
unfavored light-quark, heavy-quark, and gluon contributions. The kaon
fit-quality results in Table~\ref{tab:haps-kaff-piff-chi2} already show
a clear improvement from NLO to NNLO, especially in the SIDIS sector.
The FF comparison provides the corresponding flavor-resolved view of
this improvement. In the light-quark and strange sectors, the NLO and
NNLO results remain broadly consistent within uncertainties over much of
the fitted range, while differences in the central values reflect the
refitting required by the NNLO description of the SIA and SIDIS data.

The comparison in Figs.~\ref{fig:piff-nlo-nnlo} and
\ref{fig:kaff-nlo-nnlo} also shows that the most visible
NLO-to-NNLO response in the quark sector occurs in the favored
antiquark fragmentation functions, in particular
\(D_{\bar d}^{\pi^+}\) and \(D_{\bar s}^{K^+}\), corresponding to
\(\bar d \to \pi^+\) and \(\bar s \to K^+\). These channels are
directly connected to the charge-separated pion and kaon multiplicities
and therefore play an important role in the flavor response of the fit
when moving from NLO to NNLO. 
The gluon FFs in Figs.~\ref{fig:piff-nlo-nnlo} and
\ref{fig:kaff-nlo-nnlo} require a more cautious interpretation than the
quark FFs. In the present SIA+SIDIS setup, without identified-hadron
production data from hadronic collisions, the gluon is constrained mainly
indirectly through timelike evolution, scaling violations, higher-order
coefficient functions, and correlations with the quark FFs. 
For this reason, we display the gluon FFs separately in the corresponding
NLO--NNLO comparison panels and visually distinguish their uncertainty bands
using opposite hatch patterns. 
The displayed gluon uncertainty therefore represents the spread of the fitted
replica ensemble within the adopted methodology, rather than a complete
estimate of all possible theoretical and methodological uncertainties. 
The NLO--NNLO comparison should consequently be viewed as a perturbative
stability check within the present fit setup.

\subsection{Comparison with MAPFF1.0}\label{sec:comparison-mapff10}

To provide a quantitative benchmark against an earlier determination based
on the superseded COMPASS isoscalar-target multiplicities, we compare the
NNLO HAPS-PiFF1.0 and HAPS-KaFF1.0 fragmentation functions with the
corresponding MAPFF1.0 results of
Ref.~\cite{AbdulKhalek:2022laj}. This comparison is particularly useful
because both determinations are based on the \textsc{MontBlanc} framework
and employ a neural-network parametrization, Monte Carlo uncertainty
propagation, NNLO timelike evolution, exact NNLO SIA coefficient functions,
and approximate NNLO SIDIS corrections. It therefore provides an external
benchmark for the stability of the extracted charged-pion and charged-kaon
FFs.

The comparison is not, however, a controlled replacement of only the
COMPASS isoscalar dataset. Several experimental and theoretical ingredients
differ between the two analyses. MAPFF1.0 used the earlier COMPASS
isoscalar-target multiplicities, a lower SIDIS cut of
$Q_{\rm cut}=2.0~\mathrm{GeV}$, and NNPDF3.1 PDFs with perturbatively
generated charm. The present analyses instead combine the COMPASS 2025
proton-target multiplicities with the COMPASS 2026 revised isoscalar-target
multiplicities, use $Q_{\rm cut}=1.5~\mathrm{GeV}$ for pions and
$Q_{\rm cut}=1.7~\mathrm{GeV}$ for kaons, and adopt the corresponding
NNPDF4.0 perturbative-charm PDFs. The combination of the modern COMPASS
input and the less restrictive SIDIS cuts also results in substantially
larger SIDIS samples.

\begin{figure*}[t]
\centering
\includegraphics[width=0.30\textwidth]{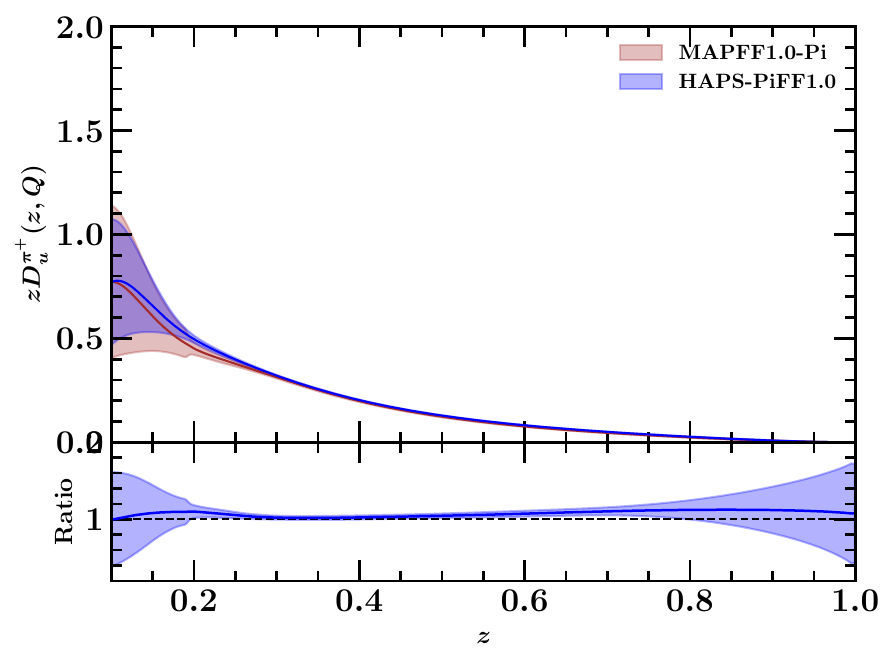}\hfill
\includegraphics[width=0.30\textwidth]{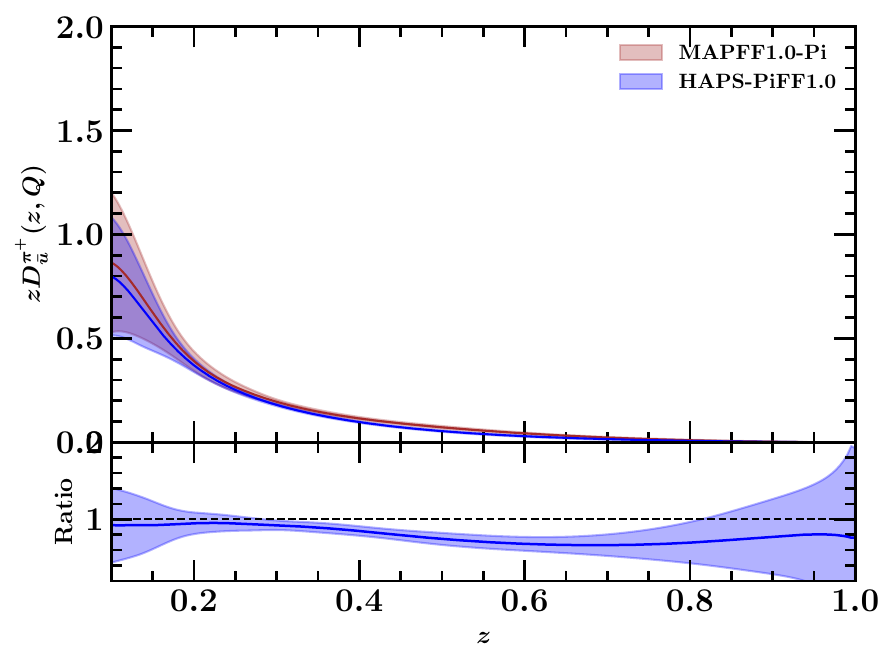}\hfill
\includegraphics[width=0.30\textwidth]{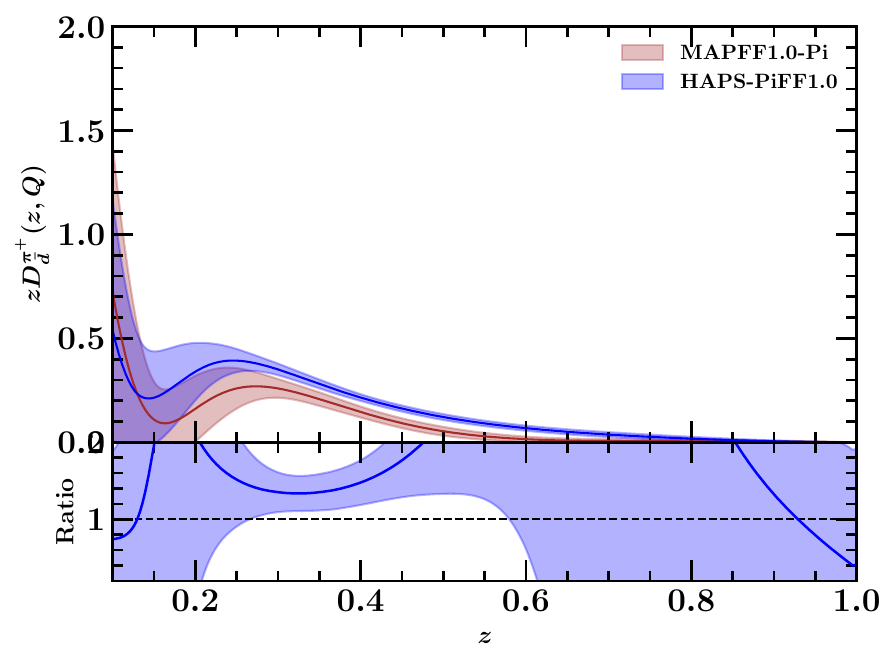}
\vspace{0.2cm}
\includegraphics[width=0.30\textwidth]{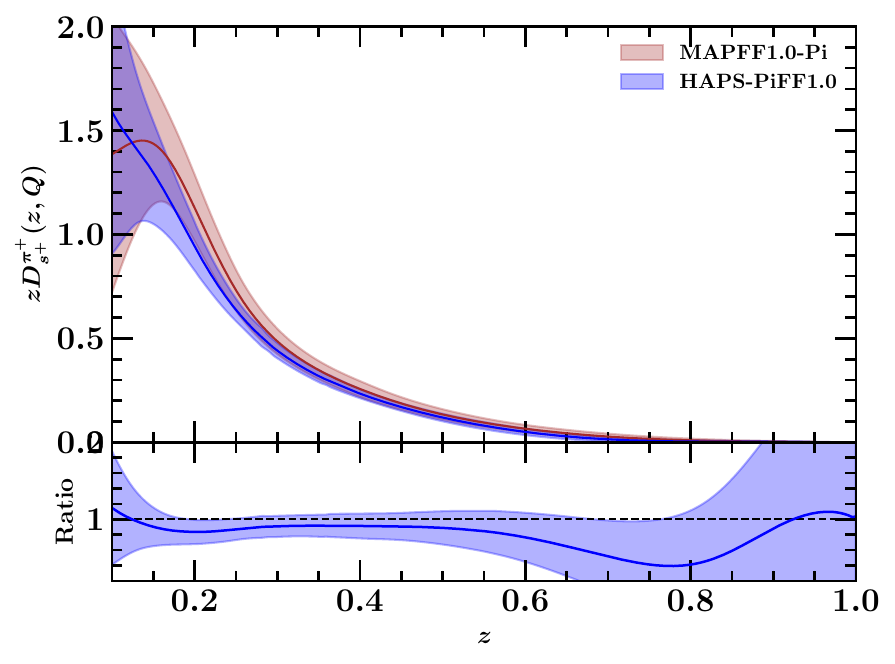}\hfill
\includegraphics[width=0.30\textwidth]{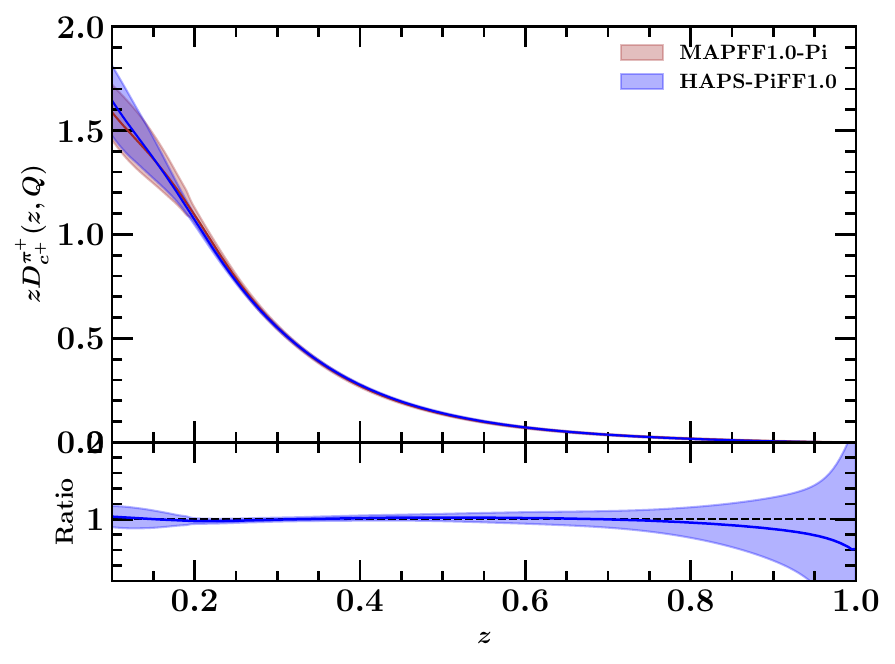}\hfill
\includegraphics[width=0.30\textwidth]{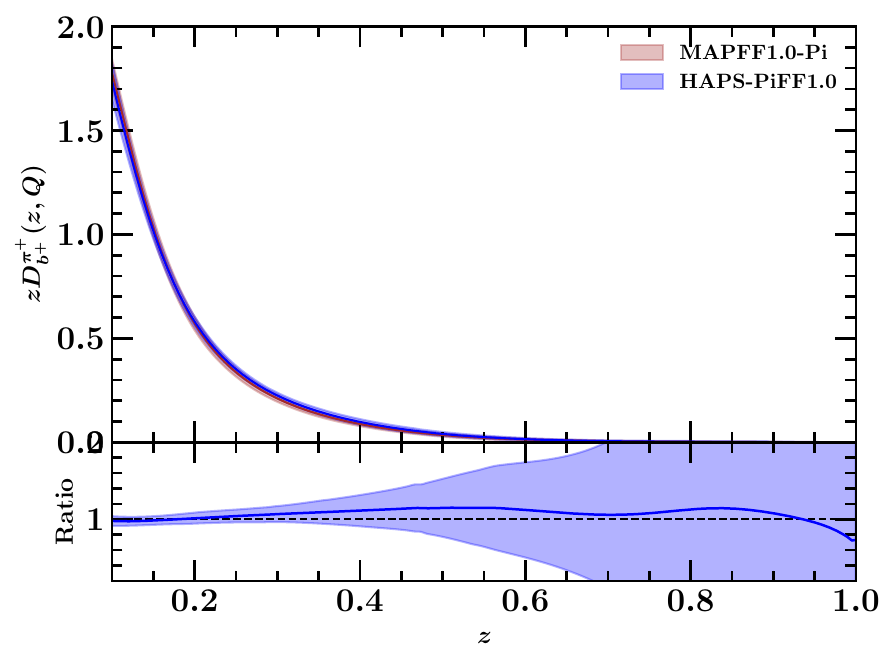}
\vspace{0.2cm}
\includegraphics[width=0.30\textwidth]{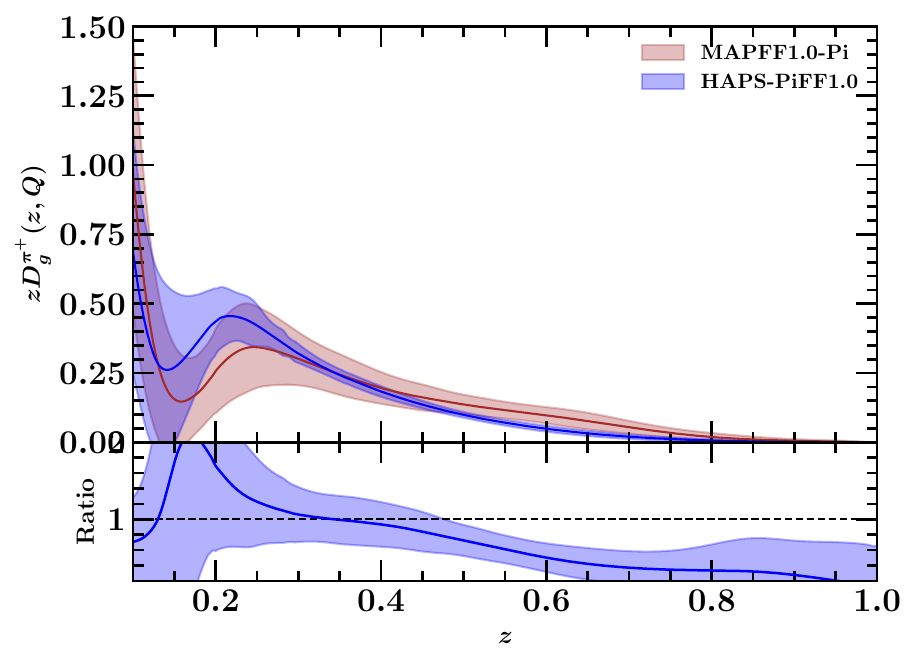}
\caption{\small \justifying
Comparison of the NNLO HAPS-PiFF1.0 and MAPFF1.0 fragmentation functions
for positively charged pions at $Q=5~\mathrm{GeV}$. The upper panels show
$zD_i^{\pi^+}(z,Q)$ for the indicated parton flavors, and the bands denote
the corresponding one-standard-deviation Monte Carlo uncertainties. The
lower panels display the direct ratio defined in
Eq.~\eqref{eq:haps-mapff-ratio}, with its uncertainty evaluated according
to Eq.~\eqref{eq:haps-mapff-ratio-uncertainty}; the dashed horizontal line
denotes unity.}
\label{fig:haps-mapff-pion} 
\end{figure*}

\begin{figure*}[t]
\centering
\includegraphics[width=0.30\textwidth]{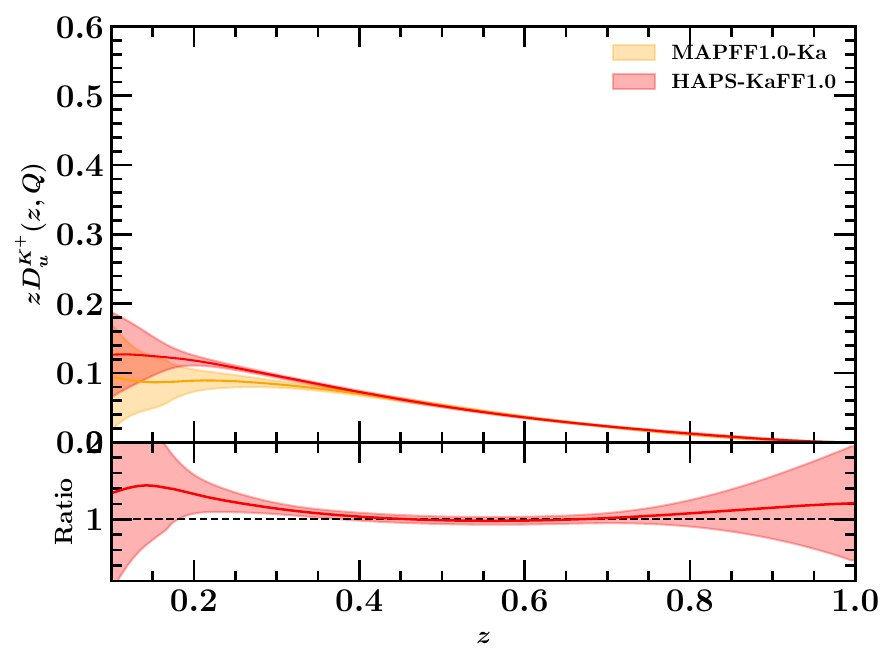}\hfill
\includegraphics[width=0.30\textwidth]{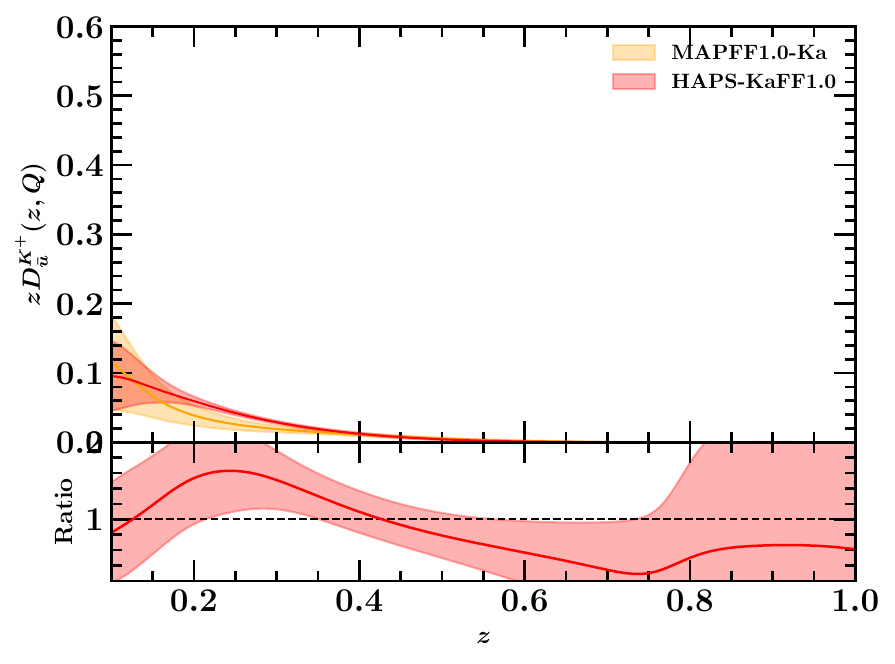}\hfill
\includegraphics[width=0.30\textwidth]{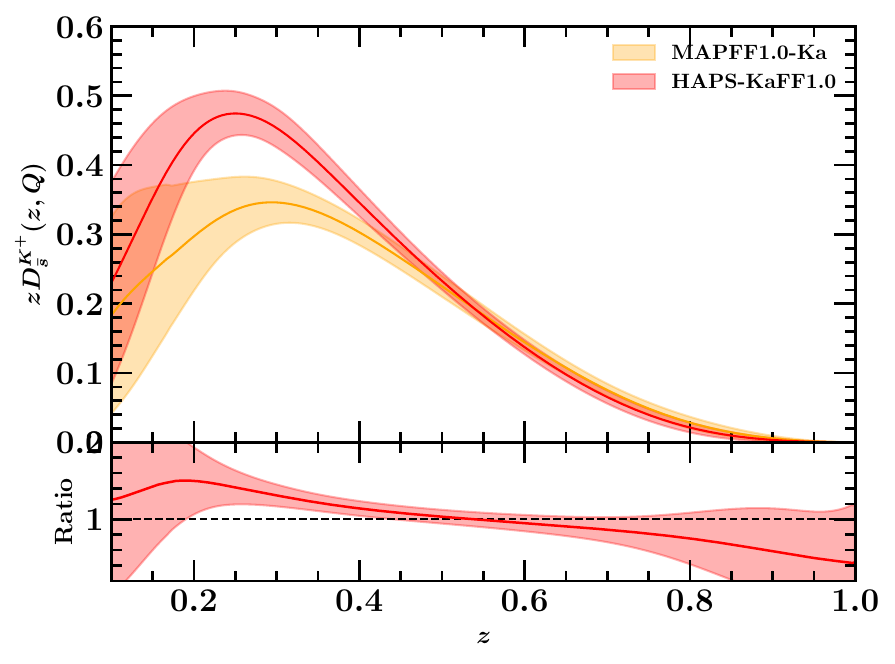}
\vspace{0.2cm}
\includegraphics[width=0.30\textwidth]{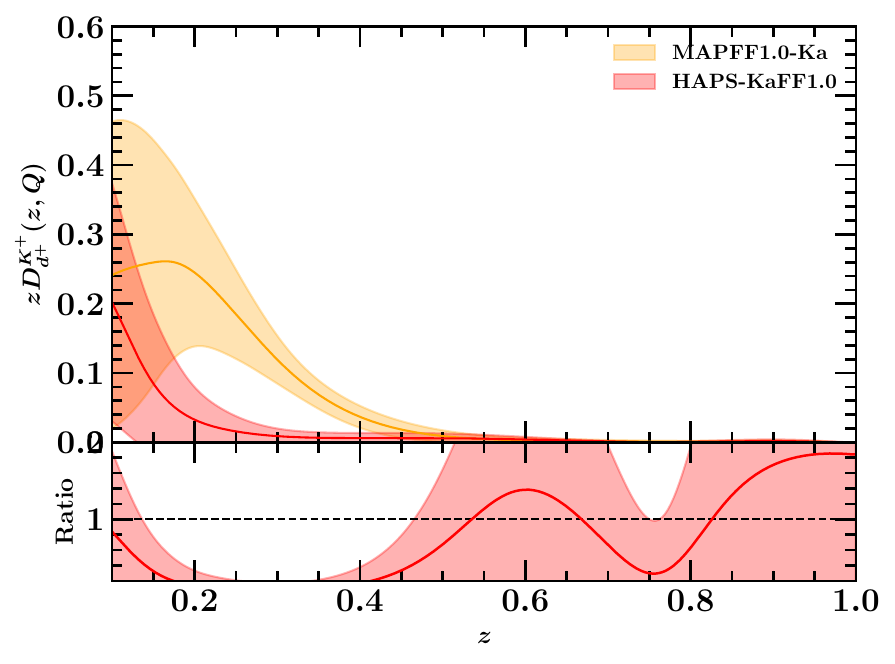}\hfill
\includegraphics[width=0.30\textwidth]{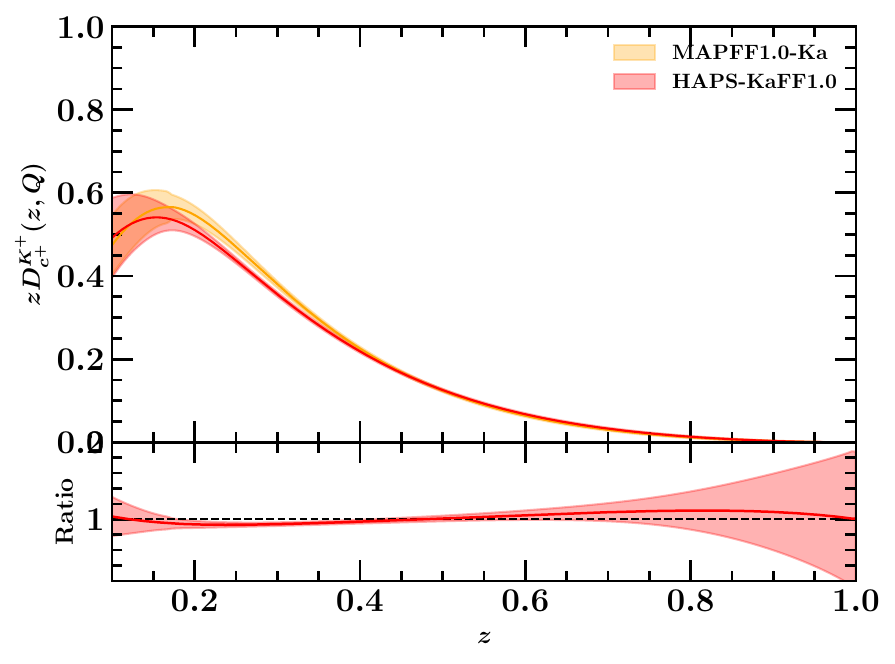}\hfill
\includegraphics[width=0.30\textwidth]{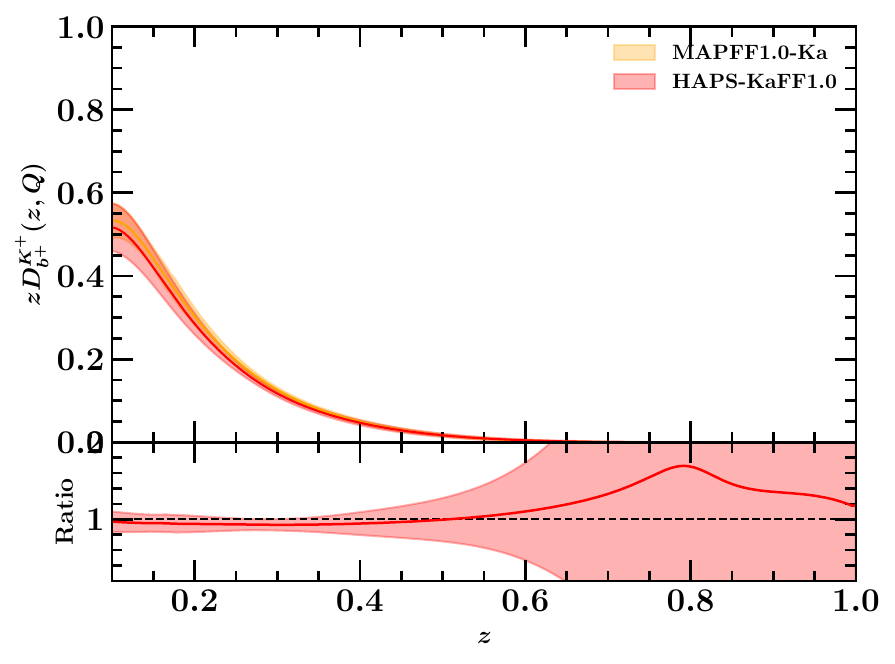}
\vspace{0.2cm}
\includegraphics[width=0.30\textwidth]{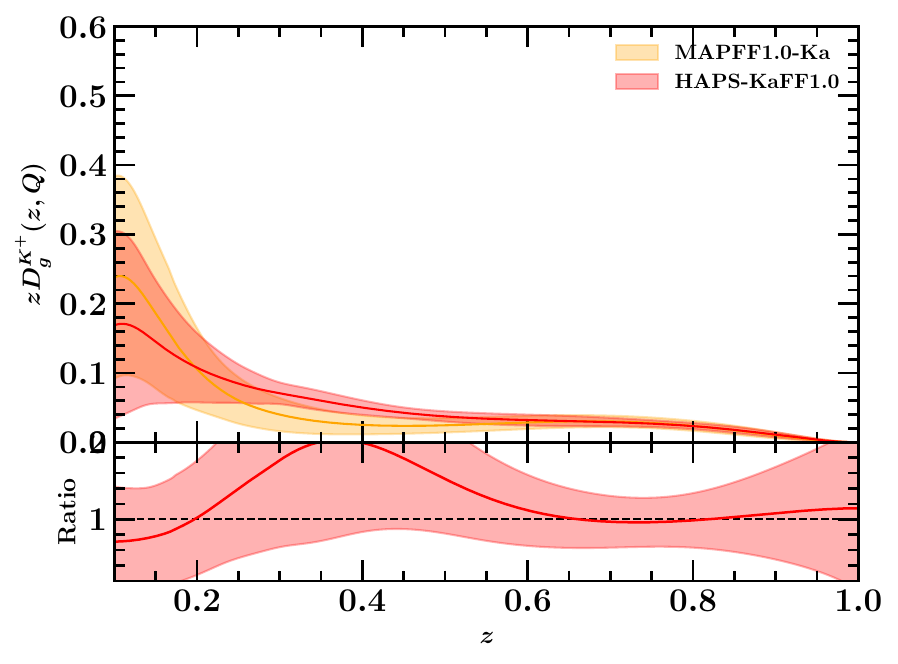}
\caption{\small \justifying
Same as Fig.~\ref{fig:haps-mapff-pion}, but for the NNLO
HAPS-KaFF1.0 and MAPFF1.0 fragmentation functions for positively charged
kaons.}
\label{fig:haps-mapff-kaon} 
\end{figure*}

Figures~\ref{fig:haps-mapff-pion} and \ref{fig:haps-mapff-kaon} compare the
resulting NNLO pion and kaon FFs at the common scale
$Q=5~\mathrm{GeV}$. The upper panels display the absolute distributions,
while the lower panels show the direct inter-set ratio
\begin{equation}
R_i^{h}(z,Q)
=
\frac{D_{i,\mathrm{HAPS1.0}}^{h}(z,Q)}
{D_{i,\mathrm{MAPFF1.0}}^{h}(z,Q)},
\qquad h=\pi^+,K^+,
\label{eq:haps-mapff-ratio}
\end{equation}
where $D_{i,\mathrm{HAPS1.0}}^{\pi^+}$ and
$D_{i,\mathrm{HAPS1.0}}^{K^+}$ denote the HAPS-PiFF1.0 and HAPS-KaFF1.0
distributions, respectively.

Because the HAPS1.0 and MAPFF1.0 fits were performed independently, no
statistical correlation between their Monte Carlo ensembles is assumed.
The one-standard-deviation uncertainty of the ratio is obtained by
first-order propagation in quadrature,

\begin{equation}
\begin{aligned}
\Delta R_i^{h}(z,Q)
&=
\left|R_i^{h}(z,Q)\right|
\Bigg[
\left(
\frac{\Delta D_{i,\mathrm{HAPS}}^{h}(z,Q)}
     {D_{i,\mathrm{HAPS}}^{h}(z,Q)}
\right)^2
\\
&\qquad\qquad+
\left(
\frac{\Delta D_{i,\mathrm{MAPFF1.0}}^{h}(z,Q)}
     {D_{i,\mathrm{MAPFF1.0}}^{h}(z,Q)}
\right)^2
\Bigg]^{1/2}.
\end{aligned}
\label{eq:haps-mapff-ratio-uncertainty}
\end{equation}
The ratio bands correspond to
$R_i^{h}\pm\Delta R_i^{h}$. This linearized expression becomes
numerically unstable when either central FF approaches zero, particularly
near $z=1$, and the corresponding endpoint ratios should therefore not be
over-interpreted.

For charged pions, the favored $u\to\pi^+$ channel is stable between the
two determinations over most of the displayed range. The central values
differ most visibly near the lower boundary, whereas at intermediate $z$
the ratio remains close to unity and the uncertainty bands overlap
substantially. The unfavored light-quark channel represented by
$D_{\bar u}^{\pi^+}=D_d^{\pi^+}$ exhibits a somewhat larger downward
displacement of the HAPS-PiFF1.0 central value over part of the
intermediate-$z$ region. Nevertheless, the two determinations remain broadly
compatible within their uncertainties over the phenomenologically relevant
range.

For the second favored channel, $\bar d\to\pi^+$,
Fig.~\ref{fig:haps-mapff-pion} shows a more visible difference than for the
favored $u$ channel. At very low $z$, the comparison is affected by large
uncertainties, while over approximately
$0.2\lesssim z\lesssim0.5$ the HAPS-PiFF1.0 central value lies above the
MAPFF1.0 result. The enhancement is most pronounced around
$z\simeq0.2$--$0.4$, after which the ratio approaches unity near
$z\simeq0.5$--$0.6$. Despite this central-value displacement, the uncertainty
bands overlap substantially and do not indicate a statistically significant
incompatibility.

For the strange combination
$D_{s^+}^{\pi^+}$, the two central values are
relatively close at low $z$. At intermediate and larger $z$, the
HAPS-PiFF1.0 central value is generally smaller than the MAPFF1.0 result.
The ratio remains moderately below unity over much of the intermediate-$z$
region and reaches its largest suppression around
$z\simeq0.7$--$0.8$. The uncertainty band is broader than for the
best-constrained favored light-quark channels and overlaps unity over a
substantial part of the displayed range.

The heavy-quark combinations $D_{c^+}^{\pi^+}$ and
$D_{b^+}^{\pi^+}$ are particularly stable. Their central values are very
similar over most of the range, with visible deviations confined mainly to
the large-$z$ endpoint, where the FFs become small and the ratios are
numerically sensitive. The pion gluon FF displays a larger dependence on
the analysis setup. Relative to MAPFF1.0, the HAPS-PiFF1.0 gluon central
value is enhanced over part of the low-to-intermediate-$z$ region and
suppressed at larger $z$. Its uncertainty is, however, substantially broader
than those of the best-constrained quark channels and overlaps unity over a
significant part of the displayed range.

Overall, the pion comparison shows broad compatibility between the two
determinations in the data-constrained region, particularly for the
$u$-quark and heavy-quark FFs. More pronounced central-value differences
occur for the favored $\bar d$ channel, the unfavored and strange sectors,
and the less directly constrained gluon FF, as well as near the boundaries
of the displayed $z$ range, where the relative uncertainties become large.

For charged kaons, Fig.~\ref{fig:haps-mapff-kaon} reveals a nonuniform
pattern across flavors. The favored $u\to K^+$ channel is broadly stable
over the intermediate-$z$ region: the two central values are close, the
ratio remains near unity, and the uncertainty bands overlap substantially.
A moderate difference is visible toward the lower-$z$ boundary. The
unfavored channel $D_{\bar u}^{K^+}=D_s^{K^+}$ displays a larger dependence
on the analysis setup, with HAPS-KaFF1.0 generally enhanced relative to
MAPFF1.0 over part of the low-to-intermediate-$z$ region. At larger $z$,
both FFs become small and the ratio becomes increasingly uncertain.

The most prominent difference in the favored sector occurs for the
$\bar s\to K^+$ channel. Around the maximum of the distribution,
approximately $z\simeq0.2$--$0.4$, the HAPS-KaFF1.0 central value lies above
the MAPFF1.0 result. The corresponding ratio is above unity in this region,
while at larger $z$ the two central values approach one another before the
ratio becomes less stable near the endpoint. The comparison therefore
indicates a visible modification of the favored strange-to-kaon FF between
the two determinations, although the uncertainty bands remain relevant when
assessing the statistical significance of this difference.

A stronger central-value difference is observed for the unfavored
$d^+$ combination. At low and intermediate $z$, MAPFF1.0 yields a larger
$d^+$ fragmentation function, whereas the HAPS-KaFF1.0 distribution
decreases more rapidly and becomes small at comparatively lower $z$. The
ratio consequently lies below unity over the region in which the MAPFF1.0
distribution remains appreciable. At larger $z$, both FFs approach zero, so
the rapidly varying ratio and its enlarged uncertainty are dominated by the
small denominator.

The heavy-quark combinations $D_{c^+}^{K^+}$ and
$D_{b^+}^{K^+}$ are considerably more stable. Their central values agree
closely over most of the intermediate-$z$ range, and the ratios remain near
unity. The visible departures at large $z$ are accompanied by increasing
uncertainties and occur where the absolute FFs are already small.

The kaon gluon FF exhibits a shape difference rather than a uniform
normalization shift. HAPS-KaFF1.0 lies below MAPFF1.0 at the lowest
displayed values of $z$, becomes moderately larger over part of the
intermediate-$z$ region, and approaches the MAPFF1.0 result again at larger
$z$. The corresponding ratio band is broad and includes unity over a
substantial part of the range. As in the pion case, the gluon is constrained
only indirectly in the present SIA+SIDIS framework through scaling
violations, timelike evolution, higher-order coefficient functions, and
correlations with the quark FFs.

Overall, the kaon comparison shows broad compatibility for the favored
$u$ channel and the heavy-quark FFs, whereas larger central-value differences
occur in the favored $\bar s\to K^+$ channel and in the unfavored
light-quark sector. The enhancement of $D_{\bar s}^{K^+}$ around its maximum
is particularly relevant because strange-to-kaon fragmentation is one of
the principal quantities constrained by the charge-separated COMPASS kaon
multiplicities.

The differences observed in both the pion and kaon comparisons reflect the
combined impact of the modern COMPASS input, the lower SIDIS scale cuts, the
updated PDF ensemble, the enlarged SIDIS datasets, and other differences in
the fitting setup. The comparison with MAPFF1.0 should therefore be regarded
as an external benchmark of the stability and flavor dependence of the
extracted FFs, rather than as a controlled one-variable study isolating the
effect of the revised COMPASS isoscalar data.

\section{Summary and Conclusions}\label{sec:Summary}

We have presented new determinations of charged-pion and charged-kaon
fragmentation functions, denoted \texttt{HAPS-PiFF1.0} and
\texttt{HAPS-KaFF1.0}, at NLO and within an NNLO perturbative setup. The latter 
combines exact NNLO timelike evolution and SIA  
coefficient functions with the approximate NNLO SIDIS corrections.
The analysis combines SIA measurements with charge-separated SIDIS
multiplicities from HERMES and COMPASS. A central feature of the present
work is the use of the modern COMPASS input, consisting of the 2025
proton-target multiplicities~\cite{COMPASS:2024gje} and the 2026 revised isoscalar-target
multiplicities from the COMPASS addendum~\cite{COMPASS:2025bfn}. 
The latter supersede the earlier COMPASS isoscalar measurements and
therefore provide the appropriate updated isoscalar companion to the new
proton-target data. 
  
The charged-kaon analysis shows a clear improvement when going from NLO
to NNLO. The global fit quality improves from
$\chi^2/N_{\rm dat}=1.13$ at NLO to $0.89$ at NNLO for
$N_{\rm dat}^{K}=1134$ fitted points. This improvement is visible in
both the SIA and SIDIS sectors.  
These results indicate that the NNLO fit provides a
more consistent description of the charged-kaon data, especially in the
SIDIS sector where the new COMPASS measurements play a central role.  
   
The charged-pion analysis exhibits a different but equally important
pattern. 
For $N_{\rm dat}^{\pi}=1295$ fitted points, the global fit quality
remains very stable, changing from
$\chi^2_{\rm tot}/N_{\rm dat}=0.8723$ at NLO to $0.8742$ at NNLO.

This global stability, however, results from compensating
changes among different data subsets.  
The pion results therefore should not be interpreted
as a uniform improvement of all datasets at NNLO, but rather as a stable
global description in which the higher-order corrections and refitting
procedure redistribute the fit quality among the SIA and SIDIS sectors.  
The proton-target and revised isoscalar-target COMPASS multiplicities
provide complementary flavor information within the present fit.

The proton-target data emphasize flavor combinations weighted by the proton
PDFs, whereas the revised isoscalar data provide a more balanced
light-flavor constraint. For charged pions, these measurements primarily
improve sensitivity to favored and unfavored light-quark fragmentation.
For charged kaons, they provide important information on the interplay
between favored light-quark fragmentation, unfavored channels, and the
strange-to-kaon sector. The extraction of the strange component should
nevertheless be interpreted with appropriate caution, since it remains
correlated with the assumed PDFs, the unfavored FFs, the imposed
kinematic cuts, and the flexibility of the nonperturbative
parametrization. 

The NLO--NNLO comparison of the resulting FFs provides a useful
diagnostic of perturbative stability. In the data-constrained region,
the pion and kaon quark FFs are generally compatible within the fitted
replica uncertainties, while larger relative differences may occur in
kinematic regions where the experimental constraints are weaker. The
gluon FF remains less directly constrained in the present SIA+SIDIS
setup. Its determination proceeds mainly through timelike evolution,
scaling violations, higher-order coefficient functions, and correlations
with the quark FFs. The displayed gluon uncertainties should therefore
be understood as the spread of the fitted replica ensemble within the
adopted methodology, rather than as a complete estimate of all possible
theoretical and methodological uncertainties.

The present analysis is complementary to previous charged-pion and
charged-kaon FF determinations, including SIA-based neural-network
extractions, MAP-style SIA+SIDIS analyses~\cite{Bertone:2017tyb,Khalek:2021gxf,AbdulKhalek:2022laj}, and 
DSS-type global fits~\cite{deFlorian:2017lwf,deFlorian:2014xna}. 
Its main distinction is the consistent inclusion of the modern COMPASS
2025 proton-target and COMPASS 2026 revised isoscalar-target
multiplicities in a common NLO and NNLO framework. The resulting
\texttt{HAPS-PiFF1.0} and \texttt{HAPS-KaFF1.0} replica sets are made
publicly available in the standard \textsc{LHAPDF} format
~\cite{HAPS-PiFF10,HAPS-KaFF10}.   

Future improvements should include a more systematic assessment of
theoretical uncertainties, including scale variations, alternative PDF
inputs, and possible power-suppressed effects in the low-scale SIDIS
region. The inclusion of additional gluon-sensitive observables, such as
identified-hadron production in hadronic collisions, would be valuable
for improving the gluon FF constraints. Further progress can also be
expected from future high-precision SIDIS measurements, in particular at
the Electron-Ion Collider (EIC)~\cite{AbdulKhalek:2021gbh,Proceedings:2026xrb,Khanpour:2026erj,Soleymaninia:2025cvi}, where 
extended kinematic coverage and 
charge-separated identified-hadron data will provide new opportunities
to test and refine the flavor structure of pion and kaon fragmentation.


\begin{acknowledgments}  
We thank the School of Particles and Accelerators at the Institute for
Research in Fundamental Sciences (IPM) for support.   
Hamzeh Khanpour appreciates the financial support from the IDUB program at AGH University of Kraków.  
Hubert Spiesberger acknowledges support by the 
Cluster of Excellence ``Precision Physics, Fundamental Interactions, 
and Structure of Matter" (PRISMA$^{++}$ EXC 2118/2) funded by the 
German Research Foundation (DFG) within the German Excellence 
Strategy (Project ID 390831469).   
\end{acknowledgments}  




\end{document}